\documentclass[journal]{IEEEtran}
\usepackage{amsmath,amsfonts}
\usepackage{algorithmic,bm}
\usepackage{array}
\usepackage{textcomp}
\usepackage{stfloats}
\usepackage{url}
\usepackage{verbatim}
\usepackage{graphicx}
\usepackage{cite}

\def\BibTeX{{\rm B\kern-.05em{\sc i\kern-.025em b}\kern-.08em
    T\kern-.1667em\lower.7ex\hbox{E}\kern-.125emX}}

\usepackage{lineno}

\usepackage{setspace}
\usepackage{amsmath}
\allowdisplaybreaks[3]
\usepackage{balance}
\usepackage{commath}
\usepackage{cancel}
\usepackage{algorithm}
\usepackage{cuted} % Required for spanning equations across columns
\usepackage{epsfig,epstopdf,graphicx,psfrag,amsmath,cases}
\usepackage{latexsym,amssymb,amsmath,epsfig,subfigure,algorithm,amsthm}
\usepackage{algorithmic}
\usepackage{color}
\usepackage{url}
\usepackage{scrtime}
\usepackage{bm}
\usepackage{graphicx}
\usepackage{bbding}
\usepackage{multicol}
\usepackage{graphicx}
\usepackage{amsthm}
\usepackage{array}
\usepackage[utf8]{inputenc}
\usepackage[english]{babel}
 \usepackage{setspace}
 \usepackage{lipsum}

\addto\captionsenglish{}
{}

\begin{document}

  \title{Full-Duplex Communication for Pinching Antenna Systems }
%  \author{
% \IEEEauthorblockN{Shaokang Hu,~\IEEEmembership{Member,~IEEE}, 
% Ruotong Zhao,~\IEEEmembership{Student Member,~IEEE}, 
% Derrick Wing Kwan Ng,~\IEEEmembership{Fellow,~IEEE}, 
% and Jinhong Yuan,~\IEEEmembership{Fellow,~IEEE}}

% \IEEEauthorblockA{School of Electrical Engineering and Telecommunications, 
% The University of New South Wales, Australia}  
% }
 \author{
\IEEEauthorblockN{Shaokang Hu, 
Ruotong Zhao, 
Derrick Wing Kwan Ng, 
and Jinhong Yuan}

\IEEEauthorblockA{School of Electrical Engineering and Telecommunications, 
The University of New South Wales, Australia}   \vspace*{-6mm}
}
\IEEEoverridecommandlockouts
\maketitle
\renewcommand{\IEEEaftertitletext}{\vspace{-0mm}}  
\begin{abstract}
This paper investigates multiuser full-duplex (FD) communication assisted by a multiwaveguide pinching-antenna system (PASS) where each dielectric waveguide can be dynamically configured to operate in either downlink transmission or uplink reception mode. We develop a channel model that accounts for in-waveguide power attenuation and cumulative pinching antenna (PA)-induced phase shift. Building on this model, we jointly optimize the waveguide modes, PA locations, PA radiation/collection coefficients, downlink beamformers, uplink receive combiners, and uplink transmit powers to maximize the downlink weighted sum-rate (WSR) subject to uplink quality-of-service (QoS) requirements and power constraints. To tackle the resulting mixed-integer nonconvex resource allocation problem, we propose a low-complexity hierarchical block coordinate descent algorithm that capitalizes on weighted minimum mean-square error, fractional programming, majorization-minimization, and the alternating direction method of multipliers (ADMM) to obtain a high-quality suboptimal solution. Simulations demonstrate that the proposed FD-PASS outperforms an FD base station equipped with fixed antennas and exhibits greater robustness against residual self-interference. Moreover, ignoring cumulative PA-induced phase shifts can produce QoS-infeasible designs.
\end{abstract}
\section{Introduction}

Sixth-generation (6G) wireless networks are expected to provide
ubiquitous, high-rate, and spectrally efficient connectivity in
dynamic propagation environments~\cite{ITURM2160}. Realizing this vision requires wireless
systems to adapt not only their transmission strategies but also their antenna configurations in response to variations in propagation conditions and user locations. However, conventional wireless systems rely
on fixed transmit and receive arrays, limiting their adaptability
mainly to electronic processing and making them particularly vulnerable
to severe path loss and blockage sensitivity, especially at high carrier
frequencies~\cite{Rangan2014mmWave}. In this regard, reconfigurable wireless architectures,
including intelligent reflecting surfaces (IRSs)~\cite{hu2021robustSecureSumRateIRS},
fluid antenna systems~\cite{new2025fluidantennas}, and movable antennas
~\cite{zhu2024movableantennas}, provide additional spatial degrees of freedom (DoFs) by dynamically adapting
the propagation environment or the physical antenna positions. However, many existing implementations of fluid and movable antennas typically provide spatial reconfigurations within regions spaced by only a few wavelengths ~\cite{new2025fluidantennas,zhu2024movableantennas}, limiting their ability to compensate for large-scale path loss or restore a blocked line-of-sight (LoS) link. Although IRSs can establish alternative  reflected paths around obstacles, the end-to-end cascaded channels generally suffer from an additional substantial path loss \cite{zhang2021active}. These limitations motivate the deployment of dynamically reconfigurable antenna
architectures that can reposition effective radiation or reception
points over an extended spatial region, thus maintaining favorable communication
links with users at different locations.
Meanwhile, full-duplex (FD) communication is an attractive technology for improving spectral efficiency by enabling simultaneous downlink transmission and uplink reception over the same frequency band. However, strong electromagnetic coupling between the transmit and receive paths gives rise to severe residual self-interference (SI), which remains a major obstacle in practical FD systems~\cite{6832464}. In fact, existing FD designs mainly mitigate SI through fixed hardware isolation and analog/digital-domain cancellation, while optimizing the transmit and receive beamformers for a fixed antenna geometry~\cite{6832464}. In contrast, reconfiguring the transmit-receive geometry introduces rich additional spatial DoFs that can be exploited to jointly enhance desired communication links and suppress SI coupling, offering greater flexibility in FD system design.

Pinching-antenna systems (PASS) enable such a reconfigurable transmit–receive geometry by guiding radio-frequency (RF) signals along dielectric waveguides and radiating or collecting them through dielectric couplers, i.e., pinching antennas (PAs), positioned in close proximity to the waveguides~\cite{Wang2025PASS,hu2026spatially,11263923}. By dynamically repositioning the PAs, PASS can adapt the effective transmit-receive geometry to the actual communication environment, thereby strengthening the desired communication links while potentially reducing transmit-receive coupling.
Moreover, owing to the interplay between guided-wave and free-space propagation characteristics of PASS, the effective channel response of PASS depends jointly on the in-waveguide propagation from the feed point to each PA and the subsequent free-space propagation between the PAs and the communication terminals. More importantly, both the amplitudes and phases of the resulting channel coefficients depend on the physical characteristics and configurations of the PAs and waveguides. Specifically, the guided signal gradually decreases along the waveguide due to inevitable practical propagation attenuation and power extraction by preceding PAs~\cite{blockage}. Additionally, each activated PA also perturbs the guided mode and introduces an additional phase
shift beyond the conventional propagation phase, which has not
been considered in existing studies, e.g.~\cite{Wang2025PASS,11263923, blockage}. When multiple PAs are activated along a transmission waveguide,
the resulting amplitude variations and cumulative phase shifts jointly determine the coherent
combination of the radiated or collected signals, and thus give rise to a trade-off between
enhancing the desired communication links and suppressing SI. Indeed, these PASS-specific channel effects directly influence how PA placement shapes the desired links and SI, motivating a closer examination of whether existing FD-PASS designs fully exploit the available reconfigurable spatial DoFs while adequately capturing its practical channel response.

Existing studies have explored FD communication in PASS from several
perspectives~\cite{Zhu2026FDWPCN,Astanjin2026FullPASS,
li2026pinching}. 
In particular, \cite{li2026pinching} focuses on single-user scenarios without addressing
multiuser FD operation or the associated interference management. Although \cite{Astanjin2026FullPASS} demonstrates
that PA placement can jointly shape both the desired and SI channels,
all these works~\cite{li2026pinching,Zhu2026FDWPCN,
Astanjin2026FullPASS} assume fixed transmit--receive waveguide assignments, thereby limiting their ability to adaptively reconfigure the PASS architecture in response to
spatial heterogeneity in the downlink and uplink user locations. Consequently, these designs cannot effectively
exploit the available spatial DoFs enabled by the reconfigurable PASS architecture.
More importantly, existing FD-PASS studies do not capture the complete channel response of PASS. In particular,
\cite{Zhu2026FDWPCN} and \cite{li2026pinching} neglect in-waveguide power
attenuation, while none of these studies accounts for the cumulative phase
shift introduced by successive PAs. Neglecting these effects may introduce
inaccuracies in the effective channel models, leading to resource allocation designs that are mismatched to underlying hardware characteristics. Such mismatches can degrade
system performance or even cause quality of service (QoS) requirements to be violated. These limitations highlight the need for a multiuser FD-PASS design under a hardware-faithful channel model that captures both in-waveguide attenuation and cumulative PA-induced phase shifts.

Motivated by these observations, we investigate a multiwaveguide multiuser
FD-PASS system with reconfigurable waveguide mode assignments and PA configurations.
Specifically, we jointly optimize the waveguide-mode assignment, PA locations and
radiation/collection coefficients, downlink beamformers, uplink receive
combiners, and uplink transmit powers to maximize the downlink weighted
sum-rate (WSR) subject to uplink QoS requirements and transmit-power constraints. To accurately characterize the hardware-dependent channel responses of PASS, we develop a configuration-dependent PASS channel model that accounts for in-waveguide power attenuation, free-space propagation, and PA-induced phase perturbations. In particular, we derive a tractable closed-form approximation for the cumulative PA-induced phase shift across multiple PAs.
Then, we develop an efficient hierarchical block coordinate descent (BCD) algorithm to tackle the resulting mixed-integer nonconvex problem. In particular, alternating direction method of multipliers (ADMM) is adopted to handle the strong coupling between the channel amplitudes and phases induced by the PA locations, enabling computationally tractable iterative solution updates.
Numerical results show that PASS improves FD performance by adapting its waveguide modes and PA positions to user locations, strengthening desired links and mitigating residual SI. Moreover, neglecting cumulative PA-induced phase shifts can lead to uplink QoS violations under the phase-aware channel model.

\emph{Notations}: Scalars, vectors, and matrices are denoted by \( x \), \( \bm{x} \), and \( \bm{X} \), respectively. 
\( \mathbb{R}^{N \times M} \) and \( \mathbb{C}^{N \times M} \) represent the sets of real and complex \( N \times M \) matrices, respectively. 
\( \mathbb{R}^{+} \) denotes the set of positive real numbers. 
The modulus of a complex scalar is \( |\cdot| \) and its conjugate is \( x^* \). The Euclidean norm of a vector is $\|\cdot\|$.
The transpose, conjugate transpose, and expectation of \( \bm{X} \) are \( \bm{X}^{\text{T}} \), \( \bm{X}^{\text{H}} \), and \( \mathbb{E}\{\bm{X}\} \), respectively. 
\( [\bm{X}]_{q,i} \) is the element at the \( q \)-th row and the  \( i \)-th column of $\bm{X}$. 
$\Re\{\cdot\}$ denotes the real part of a complex number. % \( j \) is the imaginary unit. 
A circularly symmetric complex Gaussian (CSCG) random variable with mean \( \mu \) and variance \( \sigma^2 \) has a distribution denoted by \( \mathcal{CN}(\mu, \sigma^2) \) and \( \sim \) denotes “distributed as”.

\section{System Model}\label{sysM}
\begin{figure}[t] 
  \centering
  \includegraphics[width=\columnwidth]{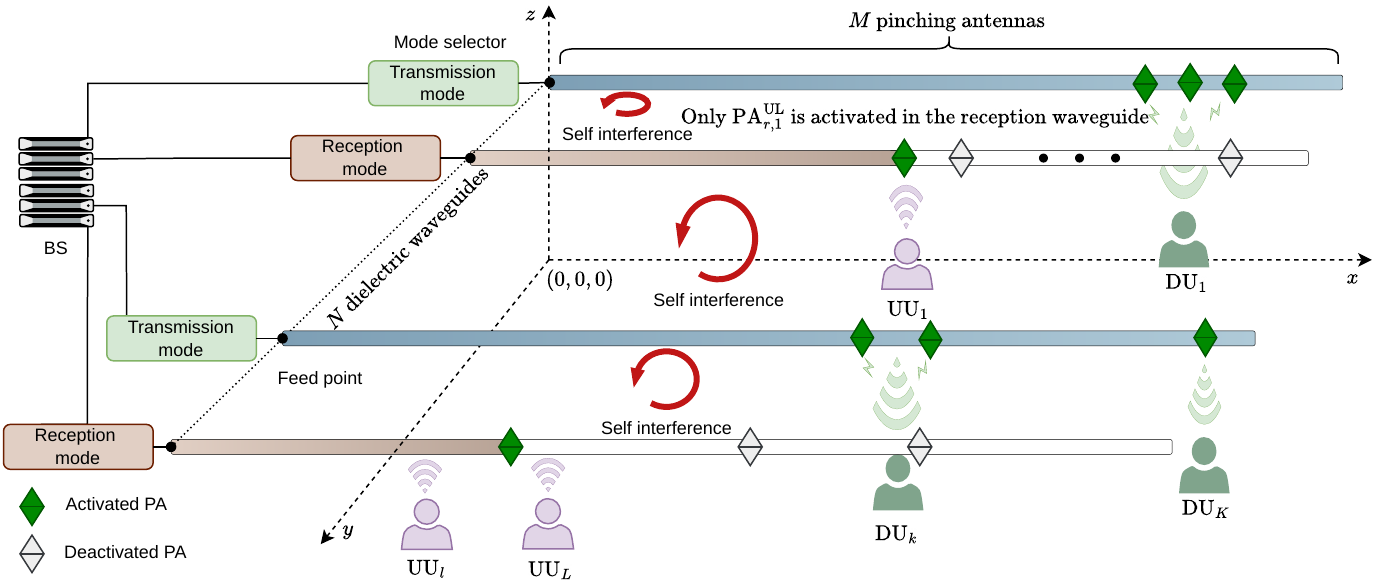}\vspace{-2mm}
  \caption{A PASS-assisted FD wireless communication system.}
  \label{fig:sys_mod}\vspace{-3mm}
\end{figure}

As shown in Fig. \ref{fig:sys_mod}, we consider a PA-assisted FD system, where a base station (BS) employs $N$ dielectric waveguides, each equipped with $M$ movable PAs that support position reconfiguration and activation/deactivation control. Unlike existing studies that assume fixed waveguide modes, e.g. \cite{Zhu2026FDWPCN,Astanjin2026FullPASS,
li2026pinching}, we optimize the mode assignment of each waveguide. Specifically, each dielectric waveguide is configured to operate in either transmission or reception mode, with its operation mode represented by the binary variable $s_n\in\{0,1\}$, $n\in\mathcal{N}$, where $s_n=0$ and $s_n=1$ denote transmission and reception modes, respectively. The associated selection mechanism will be provided in Section~\ref{sec:performanceMetric}.  We denote the numbers of transmission and reception waveguides by $N_t$ and $N_r$, respectively, such that $N=N_t+N_r$. 
The $N_t$ transmission waveguides carry distinct baseband signals for simultaneous transmission to $K$ single-antenna downlink users. All $M$ PAs on each transmission waveguide are activated for transmission.  The $N_r$ reception waveguides collect uplink signals transmitted from $L$ single-antenna uplink users. On each reception waveguide, only the first PA is activated for signal collection, whereas the remaining PAs are deactivated to prevent undesired re-radiation. %\footnote{Since a PA passively couples free-space signals into the waveguide, activating multiple PAs on the same reception waveguide may cause signals collected by the PAs further from the feed point to be reradiated by the activated PAs closer to the feed point as the collected signals propagate towards the feed point\cite{ouyang2025uplink}.}.
Assuming negligible interaction among well-separated PAs, activated transmission PAs radiate signals independently, while deactivated PAs are electromagnetically transparent and cause no additional coupling or phase perturbation \cite{johnson2002perturbation,okamoto2021fundamentals}.

We adopt a three-dimensional (3D) Cartesian coordinate system in which all $N$ dielectric waveguides are positioned at a height $d$ above the $x$-$y$ plane and aligned along the $x$-axis \cite{suzuki2022pinching,ding2024flexible}. 
We refer to the starting points of transmission waveguides as feed points, whose locations are denoted by $\bm{\psi}^{\text{DL,FP}}_t = [0, D_t^{\text{DL,FP}}, d]^{\text{T}}, \forall t  \in \mathcal{N}_t=\{1,\cdots,N_t\}$.  PA$^{\text{DL}}_{t,m}$ denotes
the $m$-th PA %\footnote{Each PA can move along a track parallel to its waveguide \cite{ding2024flexible}. We assume that PAs can be densely deployed and positioned freely along the waveguides.} 
on the $t$-th transmission waveguide, with its position denoted by $\bm{\psi}^{\text{DL,Pin}}_{t,m} = [x^{\text{DL,Pin}}_{t,m}, D_{t}^{\text{DL,FP}}, d]^{\text{T}},\forall t \in \mathcal{N}_t, \forall m \in \mathcal{M}=\{1,\ldots,M\}$. 
Similarly, the feed point of the $r$-th reception waveguide is located at $\bm{\psi}^{\text{UL,FP}}_r = [0, D_r^{\text{UL,FP}}, d]^{\text{T}}, \forall r  \in \mathcal{N}_r=\{1,\cdots,N_r\}$. PA$^{\text{UL}}_{r,m}$ denotes the $m$-th PA on the $r$-th reception waveguide with a location of $\bm{\psi}^{\text{UL,Pin}}_{r,m} = [x^{\text{UL,Pin}}_{r,m}, D_{r}^{\text{UL,FP}}, d]^{\text{T}},\forall r,m$.
We assume that the downlink and uplink users are randomly distributed across the $x$-$y$ plane. The locations of the $k$-th downlink user (DU$_{k}$) and $l$-th uplink user (UU$_l$) are denoted by
$\bm{\psi}_{k}^{\mathrm{DU}}=[x_{k}^{\mathrm{DU}},y_{k}^{\mathrm{DU}},0]^{\mathrm{T}}$
and
$\bm{\psi}_{l}^{\mathrm{UU}}=[x_{l}^{\mathrm{UU}},y_{l}^{\mathrm{UU}},0]^{\mathrm{T}}$, respectively.

\section{In-waveguide Propagation Model}
\subsection{Power Radiation Model}\label{sec:prm}
\begin{figure}[t]
  \centering
  \subfigure[Power radiation mechanism for downlink transmission. \label{fig:inwave_mod}]{%
    \includegraphics[width=\columnwidth]{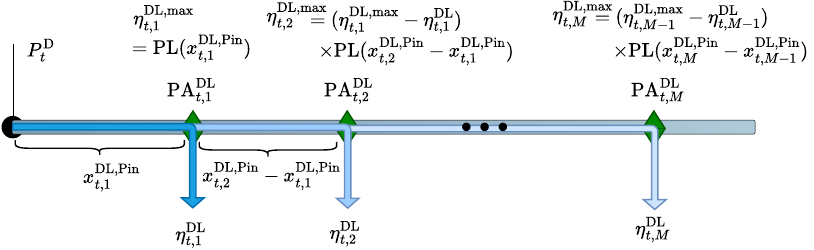}
  }\hfill \vspace{-1mm}
  \subfigure[Power collection mechanism for uplink reception. \label{fig:inwave_mod_up}]{%
    \includegraphics[width=\columnwidth]{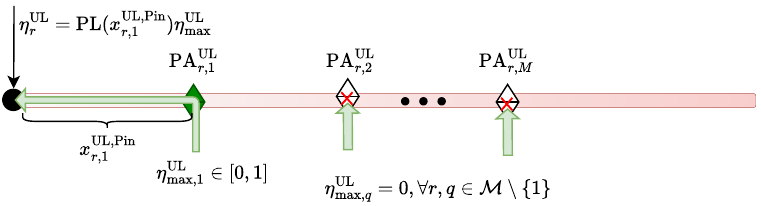}
  }
  \caption{Illustration of the downlink power radiation and uplink power collection mechanisms in the considered FD-PASS system.}
\label{fig:phase-coupling}\vspace{-5mm}
\end{figure}

Accurately accounting for in-waveguide power attenuation is essential for evaluating and optimizing the performance of PA-assisted communication systems. This paper considers a practical lossy dielectric waveguide \cite{balanis2012advanced} and adopts the in-waveguide power attenuation model in \cite{11591116, blockage}. Thus, the normalized power transmission  factor over a propagation distance $\Delta x$ is given by
\begin{align}\label{eq:Pl}
\text{PL}(\Delta x)& = e^{-2\alpha_{\text{D}}\Delta x},
\end{align} where $\alpha_{\text{D}} = \pi f_c n_{\text{eff}} \tfrac{\tan \delta_e}{c} \in \mathbb{R}^+,n_{\text{eff}} \in \mathbb{R}^+,$ and $\tan \delta_e\in \mathbb{R}^+$  denote the amplitude attenuation constant, effective refractive index of the waveguide, and effective dielectric loss tangent, respectively. Here, \(c\) and \(f_c\) denote the speed of light and signal carrier frequency, respectively.
\subsubsection{Downlink Power Model}
As illustrated in Fig.~\ref{fig:inwave_mod}, $\eta_{t,m}^{\mathrm{DL}}\in[0,1]$
denotes the effective power-radiation coefficient of
$\mathrm{PA}_{t,m}^{\mathrm{DL}}$, defined as the fraction of the power
$P_t^{\mathrm{D}}$ launched into the $t$-th transmission waveguide that is radiated by this PA, accordingly.
Hence, the radiated power $P_{t,m}^{\mathrm{rad}}$ is given by
 \begin{align}
P^{\text{rad}}_{t,m} = \eta^{\text{DL}}_{t,m}P^{\text{D}}_{t}, \qquad \forall t\in \mathcal{N}_{t}, \forall m \in \mathcal{M}.
\end{align}
Accordingly, the corresponding maximum radiated power and maximum power ratio for PA$_{t,m}$ are given by
\begin{align}\label{eq:rnm_max}
P^{\text{rad,max}}_{t,m} &= \eta^{\text{DL},\max}_{t,m}P^{\text{D}}_{t}, \qquad \forall t,m, \text{ and }\\
\eta_{t,m}^{\text{DL},\max}
&= \text{PL}(x^{\text{DL,Pin}}_{t,m})
\hspace{-1mm}-\hspace{-1mm} \sum_{i=1}^{m-1} \eta^\text{DL}_{t,i}\,\text{PL}\big(x^{\text{DL,Pin}}_{t,m} - x^{\text{DL,Pin}}_{t,i}\big),
 \forall t,m,\notag
\end{align}
 respectively. Without loss of generality, the PAs on each transmission waveguide are ordered by increasing
$x$-coordinate:
$0\leq x^{\text{Pin}}_{t,1}\leq\cdots\leq x^{\text{Pin}}_{t,M}$.
Thus, the maximum radiated power of each PA depends on its position along the waveguide and the power extracted by the preceding PAs\cite{11591116, blockage}.

\subsubsection{Uplink Power Model}
%Fig. \ref{fig:inwave_mod_up} shows the power collection mechanism for reception waveguides. Specifically,
As shown in Fig. \ref{fig:inwave_mod_up}, the maximum uplink power-collection coefficient is denoted by $\eta^\text{UL}_{\max}\in [0,1]$, which is a predefined parameter. It represents the maximum fraction of the incident free-space power that can be coupled into the $r$-th reception waveguide through  PA$^{\text{UL}}_{r,m}$. Since only the first PA on each reception waveguide is activated, the effective power collected at the feed point of the $r$-th reception waveguide is denoted as $\eta^\text{UL}_{r}=\text{PL}(x^{\text{UL, Pin}}_{r,1})\eta^\text{UL}_{\max} $, which accounts for both the PA collection ratio and the in-waveguide propagation attenuation.

\subsection{Phase-shift Model}\label{sec:phaseshift}
The in-waveguide phase response consists of two components: the intrinsic propagation phase accumulated along the dielectric waveguide and the additional phase perturbations introduced by the activated PAs. We develop a phase-shift model to characterize their combined effects.

\begin{figure}[t]
  \centering
  \subfigure[Transmission waveguide in downlink communication.\label{fig:downlink-phase}]{%
    \includegraphics[width=0.8\columnwidth]{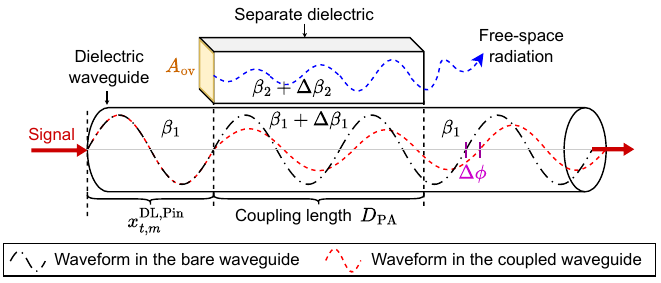}
  }\hfill %\vspace{-3mm}
  \subfigure[Reception waveguide in uplink communication.\label{fig:uplink-phase}]{%
    \includegraphics[width=0.8\columnwidth]{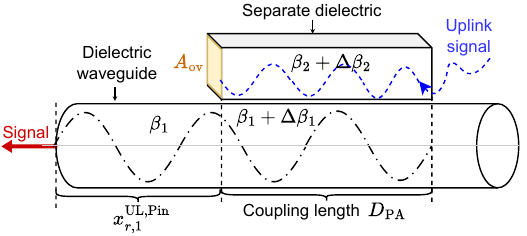}
  }%\vspace{-5mm}
  \caption{Schematic illustration of the phase shift along the dielectric waveguide caused by the waveguide coupler and PA.}
\label{fig:phase-coupling}\vspace{-5mm}
\end{figure}

\subsubsection{Downlink Phase Model}
As illustrated in Fig.~\ref{fig:downlink-phase}, since the PAs are positioned along the waveguide at nonzero distances from their feed points, 
the guided in-waveguide signals inevitably acquire propagation-induced phase shifts before reaching each 
PA~\cite{ding2024flexible}. Specifically, to account for the phase variation along the waveguide, we introduce 
the phase term at the $m$-th PA on the $t$-th transmission waveguide as
\begin{align}
\phi(x^{\text{DL,Pin}}_{t,m}) = \beta_1 x^{\text{DL,Pin}}_{t,m} \text{ and }\beta_1 = \tfrac{2\pi n_{\text{eff}}}{\lambda_0},
\end{align}
which represents the accumulated phase shift as the signal propagates from the feed point to 
the PA. Here, $\beta_1>0$ denotes the propagation constant of the uncoupled, or bare, waveguide, and $\lambda_0 = c/f_c$ is the free-space wavelength~\cite{okamoto2021fundamentals}.

When a PA is placed in close proximity to or pinched against the main dielectric
waveguide, it perturbs the local material boundary and consequently
modifies both the guided field and its propagation characteristics
\cite{johnson2002perturbation}. Thus, within the localized coupling region,
the main waveguide and the PA should be regarded as a coupled structure,
rather than as two isolated components characterized by their respective uncoupled
propagation constants $\beta_1$ and $\beta_2$ (corresponding to the bare PA). For analytical tractability, we adopt a weak-perturbation model in which the
through-propagation mode of the main waveguide experiences a local shift $\Delta\beta_1$ in its effective propagation constant\cite{johnson2002perturbation,okamoto2021fundamentals}, as shown in Fig.~\ref{fig:downlink-phase}, corresponding to an incremental phase
shift $\Delta\phi$. 
Outside the coupling region, the propagation constant of the main waveguide reverts to $\beta_1$~\cite{johnson2002perturbation}, while the phase shift accumulated within the coupling region remains as a persistent phase offset. Under the weak-perturbation assumption, the phase shifts induced by multiple activated PAs add approximately linearly along the feeder. By assuming the phase-matched case, i.e., $\beta_{1} = \beta_2$, the additional self-coupling phase shift by PA$^{\text{DL}}_{t,m}$ is approximated as a local perturbation 
\begin{align} \label{eq:add_phase}
 \Delta \phi^{\text{DL}}_{t,m}\approx -\tfrac{\pi\Delta\epsilon_r}{\lambda_0 n_{\text{eff}}}
\tfrac{A^{\text{DL}}_{\text{ov,t},m}}{A^{\text{DL}}_{\text{eff,t},m}}D_{\text{PA}^{\text{DL}}_{t,m}},
\end{align}
where $D_{\text{PA}^{\text{DL}}_{t,m}}$ denotes the constant fabricated length of PA$^{\text{DL}}_{t,m}$.   $A^{\text{DL}}_{\text{eff,t},m}$ is the modal area and  $A^{\text{DL}}_{\text{ov,t},m}$ is the area of the PA$^{\text{DL}}_{t,m}$ footprint overlapping the field \cite{snyder1983optical}. The term $\Delta\epsilon_r$ represents the variation in relative permittivity
of the system caused by introducing a PA near the dielectric waveguide. The derivation of \eqref{eq:add_phase} is provided in the appendix. 
It is worth mentioning that modeling \(\Delta\phi\) is essential for resource allocation, as pinching beamforming exploits the relative phase of the signals radiated by different PAs to enhance desired-signal combining and suppress residual SI. Indeed, neglecting $\Delta\phi$ can cause cumulative channel mismatch that worsens as the number of PAs increases, potentially diminishing the gains from PA-position optimization and compromising QoS.

%Although the PA-induced phase shifts are determined by the individual PA coupling configurations, as shown in \eqref{eq:add_phase}, we approximate them by a common effective phase shift $\Delta\phi$ for analytical tractability. 
Since a PA-induced phase shift $\Delta\phi$ introduces a
multiplicative factor of $e^{j\Delta\phi}$, the phase response
of the signal radiated by the $m$-th activated PA is
\begin{align}
g_{t,m}^{\mathrm{DL}}
=
e^{
-j(
\phi(x_{t,m}^{\mathrm{DL,Pin}})
-\sum_{q=1}^{m-1}\Delta \phi^{\text{DL}}_{t,q}
)},
\quad \forall t,m.\label{eq:phase-response}
\end{align}
Here, $\phi(x_{t,m}^{\text{DL,Pin}})$ denotes the
propagation phase accumulated from the feed point to
the $m$-th PA, while
$\sum_{q=1}^{m-1}\Delta\phi_{t,q}^{\text{DL}}$
represents the cumulative phase shift induced by
preceding activated PAs. %\footnote{The PA coordinates $x^{\text{DL,Pin}}_{t,m}$ and $x^{\text{UL,Pin}}_{r,m}$ serve as effective phase-reference points at the entrances to their respective coupling regions. The phase accumulated within each region is absorbed into these points without loss Of generality, allowing the model to capture each PA's net phase avoid describing the internal field evolution.}.

\subsubsection{Uplink Phase Model}
As shown in Fig.~\ref{fig:uplink-phase}, only the first PA on each reception waveguide is activated for signal collection. Therefore, the phase response of the signal propagating from  $\text{PA}^{\text{UL}}_{r,1}$ to the feed point of the $r$-th reception waveguide is modeled as\vspace{-1mm}
\begin{align}
g^\text{UL}_{r,1} = e^{-j\big(\phi(x^{\text{UL,Pin}}_{r,1})\big)}, \quad \forall r.%\\[-9mm] \notag
\end{align}
\section{Channel Models}

\subsection{Downlink Channel Model}
As PAs can be dynamically positioned in close proximity to users to establish near-field communications
\cite{ding2024flexible,suzuki2022pinching}, we adopt the spherical wave channel model \cite{9738442}. The free-space propagation channel from the PAs on the $t$-th transmission waveguide to DU$_k$, is given by\begin{equation}\label{eq:h_dl_k_t}
\tilde{\bm{h}}^{\text{DL}}_{k,t} \hspace{-1mm}=\hspace{-1mm}\Big[\tfrac{\sqrt{\rho_0} e^{-j \beta_0 \|\bm{\psi}^{\text{DU}}_{k} - \bm{\psi}^{\text{DL,Pin}}_{t,1}\|}}{\|\bm{\psi}^{\text{DU}}_{k} - \bm{\psi}^{\text{DL,Pin}}_{t,1}\|}  \cdots \tfrac{\sqrt{\rho_0} e^{-j\beta_0 \|\bm{\psi}^{\text{DU}}_{k} - \bm{\psi}^{\text{DL,Pin}}_{t,M}\|}}{\|\bm{\psi}^{\text{DU}}_{k} - \bm{\psi}^{\text{DL,Pin}}_{t,M}\|} \Big]^{\text{T}},
\end{equation}where $\beta_0 = \tfrac{2\pi}{\lambda_0}$ is the free-space propagation constant, $\tilde{\bm{h}}^{\text{DL}}_{k,t}\in \mathbb{C}^{M\times1}$, and \(\rho_0 = (\tfrac{c}{4\pi f_c})^2\) is the reference channel power gain at a distance of $1$ m.  Hence, the end-to-end effective downlink channel between all transmission waveguide feed points and DU$_k$ is defined as the conjugate of the physical propagation channel to facilitate beamforming design, which is given by\begin{equation}\label{eq:h_dl_k}
\hspace{-2mm}\bm{h}^{\text{DL}}_{k} \hspace{-1mm}=\hspace{-1mm}\begin{bmatrix}(\tilde{\bm{h}}^{\text{DL}}_{k,1})^{\text{T}}\text{diag}(\bm{\eta}^\text{DL}_1)\bm{g}^\text{DL}_1  \cdots(\tilde{\bm{h}}^{\text{DL}}_{k,{N_t}})^{\text{T}}\text{diag}(\bm{\eta}^\text{DL}_{N_t})\bm{g}^\text{DL}_{N_t} \end{bmatrix}^\text{H}\hspace{-2mm}.
\end{equation}
Here,  $\bm{\eta}^\text{DL}_{t} = [\sqrt{\eta^\text{DL}_{t,1}} \cdots\sqrt{\eta^\text{DL}_{t,M}} ]^{\text{T}}$  denotes the amplitude radiation vector for the $t$-th transmission waveguide, while $\bm{g}^\text{DL}_{t} = [g^\text{DL}_{t,1}\cdots g^\text{DL}_{t,M}  ]^{\text{T}}$  denotes the corresponding in-waveguide phase-response vector.

\subsection{Uplink Channel Model}
As described in Section~\ref{sysM}, only $\text{PA}^{\text{UL}}_{r,1}$ is
activated on the $r$-th reception waveguide, $r\in\mathcal{N}_r$. Thus, the link from UU$_l$ to the $r$-th reception waveguide is modeled as
\begin{align}
{h}^{\text{UL}}_{r,l}&=\tfrac{\sqrt{\rho_0} e^{- j\beta_0\|\bm{\psi}^{\text{UU}}_{l} - \bm{\psi}^{\text{UL,Pin}}_{r,1}\|}}{\|\bm{\psi}^{\text{UU}}_{l} - \bm{\psi}^{\text{UL,Pin}}_{r,1}\|}, \forall r,l.
\end{align}
The effective uplink channel vector between UU$_l$ and the BS is given by 
\begin{align}
\bm{h}^{\text{UL}}_{l}
&=
[
\sqrt{\eta^{\text{UL}}_{1}}g^\text{UL}_{1,1}{h}^{\text{UL}}_{1,l}
 \cdots 
\sqrt{\eta^\text{UL}_{N_r}}g^\text{UL}_{N_r,1}{h}^{\text{UL}}_{N_r,l}
]^{\text{T}}
,
\forall l.
\end{align}
The channel between UU$_l$ and DU$_k$ is modeled as a free-space line-of-sight (LoS) link \cite{goldsmith2005wireless}, i.e.,
$
h_{k,l}^{\text{UD}} =
\tfrac{\sqrt{\rho_0}}{ d_{k,l}^{\text{UD}}}
e^{-j\beta_0 d_{k,l}^{\text{UD}}}, \forall k,l,
$
where $d_{k,l}^{\text{UD}}$ is the distance between UU$_l$ and DU$_k$.

Furthermore, the self-interference (SI) channel matrix is denoted as $\bm{H}^{\text{SI}} \in \mathbb{C}^{N_r\times N_t}$, which is given by
%\begin{align}
%\bm{H}^{\text{SI}}&=\begin{bmatrix}\sqrt{\eta^\text{UL}_{1}}g^\text{UL}_{1,1}(\bm{h}^{\text{SI}}_1)^{\text{T}} & \cdots & \sqrt{\eta^\text{UL}_{r}}g^\text{UL}_{r,1}(\bm{h}^{\text{SI}}_r)^{\text{T}}& \cdots&\sqrt{\eta^\text{UL}_{N_r}} g^\text{UL}_{N_r,1}(\bm{h}^{\text{SI}}_{N_r})^{\text{T}}\end{bmatrix}^{\text{T}},
%\end{align}
\begin{align}
\bm{H}^{\text{SI}}&=\begin{bmatrix}\sqrt{\eta^\text{UL}_{1}}g^\text{UL}_{1,1}\bm{h}^{\text{SI}}_1 &\hspace{-2mm} \cdots \hspace{-2mm}&\sqrt{\eta^\text{UL}_{N_r}} g^\text{UL}_{N_r,1}\bm{h}^{\text{SI}}_{N_r}\end{bmatrix}^{\text{T}},
\end{align}
where
$
\bm{h}^{\text{SI}}_{r}\!\! = \![
(\bm{h}^{\text{SI}}_{1,r})^{\text{T}}\text{diag}(\bm{\eta}^{\text{DL}}_{1})\bm{g}^{\text{DL}}_1
 \cdots 
 (\bm{h}^{\text{SI}}_{N_t,r})^{\text{T}}\text{diag}(\bm{\eta}^{\text{DL}}_{N_t})\bm{g}^{\text{DL}}_{N_t}
]^{\text{T}}$ and \\ $
  \bm{h}^{\text{SI}}_{t,r}= \Big[\tfrac{\sqrt{\rho_0} e^{-j\beta_0\|\bm{\psi}^{\text{DL,Pin}}_{t,1} - \bm{\psi}^{\text{UL,Pin}}_{r,1}\|}}{\|\bm{\psi}^{\text{DL,Pin}}_{t,1} - \bm{\psi}^{\text{UL,Pin}}_{r,1}\|} \cdots \tfrac{\sqrt{\rho_0} e^{-j\beta_0\|\bm{\psi}^{\text{DL,Pin}}_{t,M} - \bm{\psi}^{\text{UL,Pin}}_{r,1}\|}}{\|\bm{\psi}^{\text{DL,Pin}}_{t,M} - \bm{\psi}^{\text{UL,Pin}}_{r,1}\|} \Big]^{\text{T}}.
$

\subsection{Signal Model}
\subsubsection{Downlink Transmission at the BS}
The downlink data signals intended for DU$_k$ are linearly precoded at baseband adopting the digital transmit beamforming vector $\bm{w}_k = [w_{k,1}, \cdots, w_{k,t}, \cdots, w_{k,N_{t}}]^{\text{T}} \in \mathbb{C}^{N_t \times 1}$, where $w_{k,t}\in\mathbb{C}$ corresponds to the transmit precoder for DU$_k$ at the $t$-th transmit waveguide.
Hence, the transmit signal $\bm{x}^{\text{D}} \in \mathbb{C}^{N_t \times 1}$ is modeled as a superposition of independent user signals, i.e.,
%\begin{align}\label{eq:s_a}
$\bm{x}^{\text{D}} = \sum_{k=1}^{K} \bm{w}_k x^{\text{D}}_k$. $P^{\text{D}}_{\max,t}$ denotes the maximum transmit power of the $t$-th transmission waveguide and the corresponding per-waveguide power constraint is $\sum_{k=1}^{K} |w_{k,t}|^2 \leq P^{\text{D}}_{\max,t}, \forall t$. Moreover, $x^{\text{D}}_k \sim \mathcal{CN}(0,1), \forall k$, with $\mathbb{E}\{|x^{\text{D}}_k|^2\} = 1$ and $\mathbb{E}\{(x^{\text{D}}_q)^* x^{\text{D}}_k\} = 0, \forall q \neq k$, represents the information-bearing symbol intended for DU$_k$.

\subsubsection{Uplink Transmission by UU$_l$}
The normalized information-bearing signal transmitted by UU$_l$ is denoted by $x^{\text{U}}_l \sim \mathcal{CN}(0,1), \forall l,$ and $p^{\text{U}}_{l}$ represents the corresponding transmit power of UU$_l$ with $P^{\text{U}}_{\max,l}$ being its maximum power budget, i.e. $0\leq p^{\text{U}}_{l}\leq P^{\text{U}}_{\max,l} $.

\subsubsection{Received Signal at DU$_k$}
The received signal at DU$_k$ is given by\begin{align}\label{eq:yDL}
{y}^{\text{DL}}_k = & (\bm{h}^{\text{DL}}_k)^{\text{H}} \bm{w}_k x^{\text{D}}_k+(\bm{h}^{\text{DL}}_k)^{\text{H}}\sum_{q=1,q\neq k}^{K} \bm{w}_q x^{\text{D}}_q+\sum_{l=1}^{L}\sqrt{p^{\text{U}}_{l}}h_{k,l}^{\text{UD}}x^{\text{U}}_l \notag\\[-2mm]&+{n}^{\text{D}}_{k}, \forall k,
\end{align}
where ${n}^{\text{D}}_{k}\sim\mathcal{CN}(0, (\sigma^{\text{D}}_{k})^2), \forall k,$ is the additive white Gaussian noise (AWGN) at DU\(_{k}\) with \((\sigma^{\text{D}}_{k})^2\) denoting its power. 
\subsubsection{Received Signal at the BS}
The BS received signal is\hspace{-1mm}
\begin{align}\label{eq:yBS}
\bm{y}^{\text{BS}} =  \sum_{l=1}^{L}\sqrt{p^{\text{U}}_{l}}\bm{h}^{\text{UL}}_l x^{\text{U}}_l+\sqrt{\rho_{\text{ANA}}}\,\bm{H}^{\text{SI}}\bm{x}^{\text{D}}+\bm{n}^{\text{BS}},
\end{align}
where $\bm{n}^{\text{BS}}\sim \mathcal{CN}(\bm{0}, (\sigma^{\text{BS}})^2\bm{I}_{N_r}),\bm{n}^{\text{BS}}\sim\mathbb{C}^{N_r\times1},$ is the thermal noise at the BS and \((\sigma^{\text{BS}})^2\) denotes the noise variance at each reception waveguide of the BS. Constant $\rho_{\text{ANA}} \in [0,1]$  captures the fraction of residual SI power after analog-domain suppression. $\rho_{\text{ANA}} =1$  and $\rho_{\text{ANA}} = 0$ correspond to no analog cancellation and perfect analog cancellation, respectively\cite{6832464}. After applying the linear receive beamformer $\bm{v}_l\in \mathbb{C}^{N_r\times 1}$ the effective received signal for $ \text{UU}_l $ is $ \bm{v}_l^\text{H}\bm{y}^{\text{BS}}$. %Here, $\bm{v}_l$ are the receive beamformers employed at the BS to recover the data signals of UU$_l$. 
\section{Problem Formulation}\label{sec:performanceMetric}
%\subsection{Achievable Data Rate} \label{sec:performanceMetric}
Since the operating modes of the dielectric waveguides are adaptively optimized according
to the channel conditions, the numbers and index sets of the transmission and
reception waveguides, as well as the dimensions of the associated variables,
may alter during the optimization. To ensure dimensional consistency, we introduce the diagonal binary
mode-selection matrix $\bm{S}=\operatorname{diag}(s_1,\ldots,s_N)$ and extend
the associated vectors and matrices to a unified $N$-dimensional representation. Under this convention, $\overline{\bm S} \triangleq {\bm I_N - \bm S}$ selects the transmission waveguides, while $\bm S$ selects the reception waveguides. Then, we redefine the location of the starting points of each waveguide as $\bm{\psi}^{\text{FP}}_n = [0, D_n^{\text{FP}}, d]^{\text{T}}, \forall n  \in \mathcal{N}=\{1,\cdots,N\}$, where $n$ indexes $N$ dielectric waveguides. 
PA$_{n,m}$ denotes the $m$-th PA on the $n$-th dielectric waveguide with a location of $\bm{\psi}^{\text{Pin}}_{n,m} = [x^{\text{Pin}}_{n,m}, D_{n}^{\text{FP}}, d]^{\text{T}}$. Specifically, the variable $\eta_{n,m} \in [0,1], \forall n,m,$ denotes the power radiation ratio or power collection ratio, depending on its corresponding operation mode of the $n$-th dielectric waveguide. 
Hence, the extended effective downlink  channel vector from the feed points of all $N$ dielectric waveguides to DU$_k,\forall k$, i.e., $\bm{h}^{\text{DL}}_{k|\text{ext}},\forall k$, is given by 
\begin{align}\label{eq:h_dl_k_ext}
\hspace{-0mm}\bm{h}^{\text{DL}}_{k|\text{ext}}\hspace{-1mm} =\hspace{-1mm}[(\tilde{\bm{h}}^{\text{DL}}_{k,1|\text{ext}})^{\text{T}}\text{diag}(\bm{\eta}_1)\bm{g}_1 {\small\cdots} (\tilde{\bm{h}}^{\text{DL}}_{k,{N}|\text{ext}})^{\text{T}}\text{diag}(\bm{\eta}_{N})\bm{g}_{N} ]^\text{H}\hspace{-2mm},\\[-9mm]\notag
\end{align}
\begin{align}
 \text{where }\tilde{\bm{h}}^{\text{DL}}_{k,n|\text{ext}} =&\Big[\tfrac{\sqrt{\rho_0} e^{-j \beta_0d^{\text{DU}}_{k,n,1}}}{d^{\text{DU}}_{k,n,1}} \cdots \tfrac{\sqrt{\rho_0} e^{-j \beta_0d^{\text{DU}}_{k,n,M}}}{d^{\text{DU}}_{k,n,M}}\Big]^{\text{T}}, \notag\\ 
\bm{\eta}_{n} = &[\sqrt{\eta_{n,1}} \cdots \sqrt{\eta_{n,M} } ]^{\text{T}}, \bm{g}_{n} = [g_{n,1} \cdots g_{n,M}] ^{\text{T}},\notag\\g_{n,m} =& e^{-j(\phi(x^{\text{Pin}}_{n,m}) - \sum_{q=1}^{m-1}\Delta \phi_{n,q})},   \\
d^{\text{DU}}_{k,n,m}=&\sqrt{(x^{\text{DU}}_{k}-x^{\text{Pin}}_{n,m})^2
+
(y^{\text{DU}}_{k}-D^{\text{FP}}_{n})^2
+
d^2}.\notag
\end{align}
The extended version of the channel from UU$_l$ to the BS, i.e., $\bm{h}^{\text{UL}}_{l|\text{ext}},\forall l$, is given by 
\begin{align}
\bm{h}^{\text{UL}}_{l|\text{ext}}&=[\sqrt{\eta_{1,1}}g_{1,1}{h}^{\text{UL}}_{1,l} \cdots
 \sqrt{\eta_{N,1}}g_{N,1}{h}^{\text{UL}}_{N,l}]^{\text{T}}\in \mathbb{C}^{N\times 1}, \label{eq:hULEXT}
\end{align}
where $d^{\text{UU}}_{l,n}
=
(
(x^{\text{UU}}_{l}-x^{\text{Pin}}_{n,1})^2
+
(y^{\text{UU}}_{l}-D^{\text{FP}}_{n})^2
+
d^2
)^{1/2}$ and ${h}^{\text{UL}}_{n,l}=\tfrac{\sqrt{\rho_0} e^{- j\beta_0d^{\text{UU}}_{l,n}}}{d^{\text{UU}}_{l,n}},
\forall l,n$. %The variable $\eta_{n,1}$ is the effective power-collection coefficient at the feed point of the reception waveguide, including the in-waveguide power attenuation.

Besides, the entry-wise $[\bm{H}^{\text{SI}}_\text{ext}]_{i,n},\forall i,n\in \mathcal{N},i\neq n$ of the extended version of SI channel $\bm{H}^{\text{SI}}_\text{ext}$ is given by
\begin{align}
 \big[\bm{H}^{\text{SI}}_{\text{ext}}\big]_{i,n}
=&
\sqrt{\eta_{i,1}}\,g_{i,1}
\big(\bm{h}^{\text{SI}}_{n,i|\text{ext}}\big)^{\text{T}}
\text{diag}(\bm{\eta}_{n})\bm{g}_n, 
\end{align}
where $d^{\text{SI}}_{i,n,m}=\sqrt{
(x^{\text{Pin}}_{n,m}-x^{\text{Pin}}_{i,1})^2
+
(D^{\text{FP}}_{n}-D^{\text{FP}}_{i})^2
}$, $\bm{h}^{\text{SI}}_{n,i|\text{ext}} =  \Big[\tfrac{\sqrt{\rho_0} e^{-j\beta_0d^{\text{SI}}_{i,n,1}}}{d^{\text{SI}}_{i,n,1}} \cdots \tfrac{\sqrt{\rho_0} e^{-j\beta_0d^{\text{SI}}_{i,n,M}}}{d^{\text{SI}}_{i,n,M}} \Big]^{\text{T}}$, and $\big[\bm{H}^{\text{SI}}_{\text{ext}}\big]_{i=n}=0$.  

%Therefore, $\gamma^{\text{DU}}_k$ and $\gamma^{\text{UU}}_l$ can be rewritten as 
%\begin{align}
%\gamma^{\text{DU}}_k = &\tfrac{\abs{(\bm{h}^{\text{DL}}_{k|\text{ext}} )^{\text{H}} (\bm{I}_{N}-\bm{S})\bm{w}_{k|\text{ext}}}^2}{\sum_{q\neq k}^{K}\abs{(\bm{h}^{\text{DL}}_{k|\text{ext}} )^{\text{H}} (\bm{I}_{N}-\bm{S})\bm{w}_{q|\text{ext}}}^2+\sum_{l=1}^{L}p^{\text{U}}_{l}\abs{h_{k,l}^{\text{UD}}}^2+(\sigma^{\text{D}}_k)^2},\forall k\text{, and}\label{eq:sinr_du}\\
%\gamma^{\text{UU}}_l = & \tfrac{p^{\text{U}}_{l}\abs{\bm{v}_{l|\text{ext}}^{\text{H}}\bm{S}\bm{h}^{\text{UL}}_{l|\text{ext}}}^2}{\sum_{q\neq l}^{L}p^{\text{U}}_{q}\abs{\bm{v}_{l|\text{ext}}^{\text{H}}\bm{S}\bm{h}^{\text{UL}}_{q|\text{ext}}}^2+
%\rho_{\text{ANA}}\sum_{k=1}^{K}\abs{\bm{v}_{l|\text{ext}}^{\text{H}}\bm{S}\bm{H}^{\text{SI}}_{\text{ext}}(\bm{I}_{N}-\bm{S})
%\bm{w}_{k|\text{ext}}}^2+(\sigma^{\text{BS}})^2\|\bm{v}_{l|\text{ext}}^{\text{H}}\bm{S}\|^2},\forall l,\label{eq:sinr_uu}
%\end{align}
%respectively.

Therefore, the achievable data rate (bit/s/Hz) between the FD BS and  DU$_k$ is given by
\begin{align}
R^{\text{DU}}_k=&\log_2(1+\gamma^{\text{DU}}_k), \forall k\text{, where}\label{eq:sinr_du}\\
\hspace{-2mm}\gamma^{\text{DU}}_k\hspace{-1mm} = &\tfrac{\abs{(\bm{h}^{\text{DL}}_{k|\text{ext}} )^{\text{H}} \overline{\bm{S}}\bm{w}_{k|\text{ext}}}^2}{\sum_{q\neq k}^{K}\abs{(\bm{h}^{\text{DL}}_{k|\text{ext}} )^{\text{H}} \overline{\bm{S}}\bm{w}_{q|\text{ext}}}^2+\hspace{0mm}\sum_{l=1}^{L}p^{\text{U}}_{l}\abs{h_{k,l}^{\text{UD}}}^2\hspace{-0mm}+(\sigma^{\text{D}}_k)^2}.\notag
\end{align}
As for uplink reception, we assume that the BS employs a linear
receiver for decoding of the received uplink information~\cite{goldsmith2005wireless}. The achievable uplink data rate of UU$_l$ is 
\begin{align}
\hspace{-1mm}R^{\text{UU}}_l\hspace{-1mm}=&\log_2(1+\gamma^{\text{UU}}_l), \forall l,\label{eq:sinr_uu}\\%\label{eq:r_uu}
\hspace{-1mm}\gamma^{\text{UU}}_l\hspace{-1mm} =& \tfrac{p^{\text{U}}_{l}\abs{\bm{v}_{l|\text{ext}}^{\text{H}}\bm{S}\bm{h}^{\text{UL}}_{l|\text{ext}}}^2}{\hspace{-2mm}\sum\limits_{\substack{ q=1, \\q\neq l}}^{L}p^{\text{U}}_{q}\abs{\bm{v}_{l|\text{ext}}^{\text{H}}\bm{S}\bm{h}^{\text{UL}}_{q|\text{ext}}}^2\hspace{-1mm}+
\rho_{\text{ANA}}\sum\limits_{k=1}^{K}\abs{\bm{v}_{l|\text{ext}}^{\text{H}}\bm{S}\bm{H}^{\text{SI}}_{\text{ext}}\overline{\bm{S}}
\bm{w}_{k|\text{ext}}}^2\hspace{-1mm}+(\sigma^{\text{BS}})^2\|\bm{v}_{l|\text{ext}}^{\text{H}}\bm{S}\|^2}.\notag
\end{align}

\subsection{Optimization Problem Formulation}
Motivated by the typical higher traffic demand in the downlink than in the uplink, we maximize the
downlink WSR by
jointly optimizing
$\{x^{\text{Pin}}_{n,m},\eta_{n,m},\bm{w}_{k|\text{ext}},
\bm{v}_{l|\text{ext}},p_l^{\mathrm{U}},s_n\}$.
The predefined weights $\mu_k\geq 0$ balance throughput and user fairness. The resulting optimization problem is formulated as
\begin{align}\label{proposed_formulation_origion}
 & \underset{  \substack{x^{\text{Pin}}_{n,m}, \eta_{n,m},s_n, \bm{w}_{k|\text{ext}},\bm{v}_{l|\text{ext}},p^{\text{U}}_l}}{\text{maximize}} \,\,  \sum_{k=1}^{K}\mu_k R^{\text{DU}}_k \\
  \text{s.t.}\,\,
 & \text{C1}\hspace{-1mm}:R^{\text{UU}}_l \geq R^{\text{UU}}_{\text{min}}, \forall l,
\notag\\[-0mm]
 & \text{C2}\hspace{-1mm}: 0\leq x^{\text{Pin}}_{n,m} \leq x^{\max},\forall n,m,\notag\\[-0mm]
  & \text{C3}\hspace{-1mm}: x^{\text{Pin}}_{n,m+1} -x^{\text{Pin}}_{n,m} \geq D_{\text{PA}}+\delta x,\forall n, m\in\mathcal{M}\setminus \{M\},
\notag\\[-0mm]
 & \text{C4}\hspace{-1mm}:0\leq \eta_{n,m}\leq (1 - s_n) \Big(\text{PL}\big(x^{\text{Pin}}_{n,m}\big) \notag\\&\hspace{6mm}-\sum_{i=1}^{m-1} \eta_{n,i}\,
  \text{PL}\big(x^{\text{Pin}}_{n,m} - x^{\text{Pin}}_{n,i}\big)\Big), \forall n , m\in\mathcal{M}\setminus \{1\}, \notag\\[-0mm] 
 & \text{C5}\hspace{-1mm}:0\leq \eta_{n,1}\leq (1 - s_n)\text{PL}\big(x^{\text{Pin}}_{n,1}\big)+s_n\eta^\text{UL}_{\max}\text{PL}(x^{\text{Pin}}_{n,1}),\forall n,\displaybreak[3]\notag\\[-0mm]
 & \text{C6}\hspace{-1mm}:\sum_{k=1}^{K} |[\overline{\bm S}\bm{w}_{k|\text{ext}}]_n|^2 \leq P^{\text{D}}_{\max,n}, \forall n, \notag\\&
 \text{C7}\hspace{-1mm}:0\le p_l^\text{U} \le P^{\text{U}}_{\max,l}, \forall l,\notag\\
 & \text{C8}\hspace{-1mm}:s_n \in \{0,1\},\forall n, \text{C9}\hspace{-1mm}:1 \leq \sum_{n=1}^{N} s_n\leq N-1.\notag
\end{align}

Constraint $\text{C1}$ enforces the minimum uplink rate requirement
$R^{\mathrm{UU}}_{\min}$ for each uplink user for QoS provisioning. Constraints $\text{C2}$ and $\text{C3}$ restrict the PA locations to
$[0,x^{\max}]$ and enforce a minimum edge-to-edge spacing of
$\delta_x$, corresponding to a reference-point separation of
$D_{\mathrm{PA}}+\delta_x$, respectively. Constraints $\text{C4}$ and
$\text{C5}$ impose mode-dependent radiation or collection coefficient limits. Note that \(\eta_{n,1}\) denotes the effective power-collection coefficient at the feed point, including in-waveguide attenuation. As constraint C5 is active at the optimum,  we have $
\eta_{n,1}=\eta_{\max}^{\mathrm{UL}}P_L(x_{n,1}^{\mathrm{Pin}})$ when the \(n\)-th waveguide is selected for reception (\(s_n=1\)). %Specifically, in transmission mode ($s_n=0$), $\eta_{n,m}$ is limited by the in-waveguide attenuation caused by preceding PAs. As in reception mode ($s_n=1$), only the first PA is activated, with its collection ratio bounded by $\eta^{\text{UL}}_{\max}$, while the remaining PAs are deactivated. 
Constraints $\text{C6}$ and $\text{C7}$ impose the per-waveguide downlink and per-user uplink power budgets,
$P^{\text{D}}_{\max,n}$ and $P^{\text{U}}_{\max,l}$, respectively. Constraint
$\text{C8}$ restricts each waveguide to either transmission or reception mode,
while $\text{C9}$ ensures that at least one waveguide is assigned to each mode.

\section{Solutions}
\subsection{Overall Algorithm Framework}
\label{solF}
\begin{figure}[t] 
  \centering
  \includegraphics[width=\columnwidth]{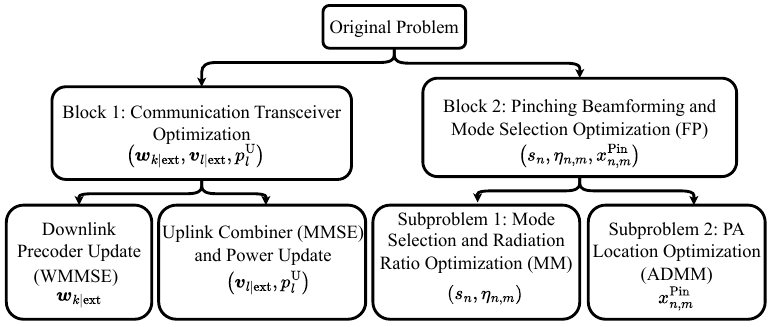}\vspace{-3mm}
  \caption{Block diagram of the proposed algorithmic framework.}\vspace{-5mm}
  \label{fig:overall_solution_frame}
\end{figure}
Problem~\eqref{proposed_formulation_origion} is a mixed-integer and non-convex optimization program owing to residual-SI coupling, geometry-dependent channel amplitudes and phases, and binary mode selection.
To address these challenges, we develop a computationally efficient hierarchical framework integrating BCD, weighted minimum mean-square error (WMMSE),
fractional programming (FP),
majorization-minimization (MM), and ADMM to obtain a high-quality suboptimal solution. As shown in Fig. \ref{fig:overall_solution_frame}, the proposed framework decomposes the problem into two blocks: Block~1 optimizes $\{\bm{w}_{k|\text{ext}},\bm{v}_{l|\text{ext}},p^{\text{U}}_l\}$, while Block~2 optimizes $\{s_n,\eta_{n,m},x^{\text{Pin}}_{n,m}\}$.

\label{sol:frame}
\subsection{Block 1: Communication Transceiver
Optimization}
With fixed $\{s_n, x^{\text{Pin}}_{n,m}, \eta_{n,m}\}$, we alternatively optimize  \(\bm{w}_{k|\text{ext}}\) and  \{\(\bm{v}_{l|\text{ext}}, p_l^{\mathrm U}\)\} until convergence.

\subsubsection{Optimizing Downlink Beamforming Vector, $\bm{w}_{k|\text{ext}}$}
For fixed $\{\bm{v}_{l|\text{ext}},p^{\text{U}}_l\}$, the downlink subproblem is
\begin{align}
  \underset{ \bm{w}_{k|\text{ext}}}{\text{maximize}} \,\,&  \sum_{k=1}^{K}\mu_k\log_2(1+\gamma^{\text{DU}}_k)\label{eq:block1_dl_problem}\\[-2mm]
  \text{s.t.}\,\,&\text{C1}\hspace{-1mm}:\rho_{\text{ANA}}\Gamma
\sum_{k=1}^{K}
\left|
\bm v_{l|\text{ext}}^{\text{H}}
\bm{S}\bm{H}^{\text{SI}}_{\text{ext}}\overline{\bm{S}}
\bm w_{k|\text{ext}}
\right|^2
\le
\tau_l,
 \forall l,  \text{C6},\notag
\end{align}
where $\tau_l
=
p_l^\text{U}
\left|
\bm v_{l|\text{ext}}^\text{H}
\bm{S}\bm{h}^{\text{UL}}_{l|\text{ext}}
\right|^2
-
\Gamma
\sum_{q\neq l}^L
p_q^\text{U}
\left|
\bm v_{l|\text{ext}}^\text{H}
\bm{S}\bm{h}^{\text{UL}}_{q|\text{ext}}
\right|^2
-
\Gamma
(\sigma^{\text{BS}})^2
\|
\bm v_{l|\text{ext}}^\text{H} \bm S
\|^2$ and $\Gamma \triangleq 2^{R^{\text{UU}}_{\text{min}}}-1$. Note that, if $\exists\tau_l<0$, discard $\{\bm v_{l|\mathrm{ext}},p_l^{\mathrm U}\}$, since C1 is infeasible for any
downlink beamformers. Then, for $\tau_l\geq0$, apply the WMMSE framework \cite{Shi2011IWMMSE}  to the following equivalent reformulation of 
\eqref{eq:block1_dl_problem}:
  \begin{align}
\min_{\bm{w}_{k|\text{ext}},\bm{u}_k,\lambda_k} &\sum_{k=1}^K \mu_k( \lambda_k \, \varepsilon_k(u_k,\bm{w}_{k|\text{ext}}) -\ln(\lambda_k))\notag \\ \text{s.t.}& \quad \text{C1},\text{C6},
\label{eq:wmmse_tx_qp_1}
\end{align}
where $u_k\in\mathbb{C}$ and $\lambda_k>0$ denote the linear receive
filter and mean square error (MSE) weight for DU$_k$, respectively. The MSE $\varepsilon_k=\mathbb{E}\big\{|u_k y_k^{\text{DL}} - x_k^{\text D}|^2\big\}$ is given by
\begin{align}
\varepsilon_k(u_k,\bm{w}_{k|\text{ext}})&=  |{u}_k|^2 \sum_{j=1}^K |(\bm{h}^{\text{DL}}_{k|\text{ext}})^{\text{H}}\overline{\bm{S}} \bm{w}_{j|\text{ext}}|^2 +1\label{eq:mse_full}\\&- 2 \Re\!\{ {u}_k (\bm{h}^{\text{DL}}_{k|\text{ext}})^{\text{H}}\overline{\bm{S}} \bm{w}_{k|\text{ext}}  \} + |{u}_k|^2 \sigma_{k,\text{Deff}}^2,\notag
\end{align}
where $\sigma_{k,\text{Deff}}^2 = \sum_{l=1}^L p_l^\text{U} |h_{k,l}^{\text{UD}}|^2 + (\sigma_k^\text{D})^2$.
A stationary point of   \eqref{eq:wmmse_tx_qp_1} can be obtained via the following three cyclic steps\cite{Shi2011IWMMSE}. In particular, each step update admits either a closed-form solution or a convex optimization problem:\\
\noindent\textbf{WMMSE step 1}: For each DU $k$, the optimal linear receive filter ${u}_k$ is given by MMSE criterion
\begin{equation}
{u}_k^\star = \tfrac{\left((\bm{h}^{\text{DL}}_{k|\text{ext}})^{\text{H}}\overline{\bm{S}} \bm{w}_{k|\text{ext}}\right)^*}{\sum_{j=1}^K \abs{(\bm{h}^{\text{DL}}_{k|\text{ext}})^{\text{H}}\overline{\bm{S}} \bm{w}_{j|\text{ext}}}^2+\sigma_{k,\text{Deff}}^2}.
\label{eq:mmse_receiver}
\end{equation}
\noindent\textbf{WMMSE step 2}: The optimal weight $\lambda_k$ can be found by 
\begin{equation}
\lambda_k^\star = \varepsilon_k({u}_k^\star, \bm{w}_{k|\text{ext}})^{-1}.
\label{eq:mse_weight}
\end{equation}
\noindent \textbf{WMMSE step 3}: Fixing $(\bm{u}_k, \lambda_k)$, \eqref{eq:wmmse_tx_qp_1} is equivalent to 
\begin{equation}
\min_{ \bm{w}_{k|\text{ext}}} \; \sum_{k=1}^K \mu_k \lambda_k \, \varepsilon_k({u}_k,  \bm{w}_{k|\text{ext}}) \quad \text{s.t.} \quad \text{C1,C6}.
\label{eq:wmmse_tx_qp}
\end{equation}

In \textbf{WMMSE Step 3}, problem \eqref{eq:wmmse_tx_qp} is a convex quadratically constrained quadratic program (QCQP), whose  Karush–Kuhn–Tucker (KKT) conditions~\cite{boyd2004convex} yield
\begin{align}
\bm{\Psi}\bm{w}_{k|\text{ext}}^\star
=&\mu_k\lambda_k u_k^{*}\overline{\bm{S}} 
\bm{h}^{\text{DL}}_{k|\text{ext}}, \text{where}
\label{eq:tx_beamformer}\\
\bm{\Psi}\triangleq &
\sum_{q=1}^{K}
\mu_q \lambda_q |u_q|^2\overline{\bm{S}} 
\bm{h}^{\text{DL}}_{q|\text{ext}}
(\bm{h}^{\text{DL}}_{q|\text{ext}})^{\text{H}}\overline{\bm{S}} 
+\sum_{l=1}^{L}\zeta_l^{\text{C1}}\bm G_l
\notag \\&
+
\overline{\bm{S}}\operatorname{diag}\!\big([\zeta^{\text{C6}}_1,\ldots,\zeta^{\text{C6}}_N]\big)\overline{\bm{S}}
\text{ and }\notag\\
\bm G_l
\triangleq&
\rho_{\text{ANA}}
\Gamma
\big(\bm S
\bm H_{\text{ext}}^{\text{SI}}
\overline{\bm{S}} \big)^{\text{H}}
\bm v_{l|\text{ext}}
\bm v_{l|\text{ext}}^{\text{H}}
\bm S
\bm H_{\text{ext}}^{\text{SI}}
\overline{\bm{S}} .\notag
\end{align}
 The nonnegative Lagrange multipliers $\zeta^{\text{C1}}_l$ and $\zeta^{\text{C6}}_n$ are associated with $\text{C1}$ and $\text{C6}$, respectively, which can be efficiently
obtained via the projected subgradient method \cite{boyd2004convex}.
% The dual variables $\zeta^{\text{C1}}_l, \forall l,$ and $\zeta^{\text{C6}}_n,  \forall n,$ can be efficiently
% obtained via the projected subgradient method:
% \begin{align}
% \zeta_{n}^{\text{C6}(t_{\text{D}}+1)}
% \hspace{-1mm}=&
% \Big[
% \zeta_{n}^{\text{C6}(t_{\text{D}})}
% \hspace{-1mm}+\hspace{-1mm}
% \alpha^{(t_{\text{D}})}
% \big(
% \sum_{k=1}^{K}
% \abs{
% [\overline{\bm S}\bm w_{k|\text{ext}}^{(t_{\text{D}})}]_n
% }^2\hspace{-1mm}
% -\hspace{-1mm}
% P_{\max,n}
% \big)
% \Big]^+\hspace{-3mm},\notag
% \\
% \zeta_{l}^{\text{C1}(t_{\text{D}}+1)}
% \hspace{-1mm}=&
% \Big[
% \zeta_{l}^{\text{C1}(t_{\text{D}})}
% \hspace{-1mm}+\hspace{-1mm}
% \alpha^{(t_{\text{D}})}
% \big(
% \sum_{k=1}^{K}
% (\bm w_{k|\text{ext}}^{(t_{\text{D}})})^{\text{H}}
% \bm G_l
% \bm w_{k|\text{ext}}^{(t_{\text{D}})}\hspace{-1mm}
% -\hspace{-1mm}
% \tau_l
% \big)
% \Big]^+. \label{eq:dual_update}
% \end{align}
% Here, $(t_{\text{D}})$ denotes the dual-variable update index,  $\alpha^{(t_{\text{D}})}>0$ denotes the step size, and
% $[x]^+ \triangleq \max\{x,0\}$.
% The above iterations are repeated until convergence.
\subsubsection{Updating Uplink Receiver Combiners, $\bm{v}_{l|\text{ext}}$}
For any $\{\bm{w}_{k|\mathrm{ext}},p_l^{\mathrm U}\}$, the combiner
update in Block~1 reduces to maximizing a generalized Rayleigh quotient \cite{goldsmith2005wireless}
for each uplink user. Accordingly, the optimal receive beamformer is
given by the MMSE combiner:
\begin{align}
\bm v_{l|\text{ext}}^\star
=&
\bm R_l^{-1}
\bm S
\bm h^{\text{UL}}_{l|\text{ext}},
\quad \forall l,\notag
\\
\bm R_l
=&
\sum_{q\neq l}
p_q^{\text U}
\bm S
\bm h^{\text{UL}}_{q|\text{ext}}
(\bm S
\bm h^{\text{UL}}_{q|\text{ext}})^{\text H}
+
\rho_{\text{ANA}}
\sum_{k=1}^{K}
\bm S
\bm H_{\text{ext}}^{\text{SI}}
\overline{\bm S} \notag\\&\times
\bm w_{k|\text{ext}}
\Big(
\bm S
\bm H_{\text{ext}}^{\text{SI}}
\overline{\bm S}
\bm w_{k|\text{ext}}
\Big)^{\text H}
+
(\sigma^{\text{BS}})^2
\bm{I}_{N} .
\label{eq:mmse_combiner}
\end{align}
Since $s_n$ is fixed in Block 1, the MMSE receiver is designed in the extended $N$-dimensional space. Hence, we adopt $(\sigma^{\text{BS}})^2
\bm{I}_{N} $ as the positive definite noise covariance matrix such that $\bm R_l$ remains positive definite and invertible.
\subsubsection{Updating Uplink Transmit Powers, $p_l^{\text U}$}
With fixed $\bm{w}_{k|\text{ext}}$ and the MMSE combiners $\bm v_{l|\text{ext}}^\star$ obtained in Step 2, by substituting \eqref{eq:mmse_combiner} into \eqref{eq:sinr_uu}, we have
\begin{equation}
\gamma_l^{\text{UU}}
=
p_l^{\text U}
(\bm S \bm h^{\text{UL}}_{l|\text{ext}})^{\text H}
\bm R_l^{-1}
\bm S \bm h^{\text{UL}}_{l|\text{ext}}.
\label{eq:sinr_after_mmse}
\end{equation}

Since the uplink rates appear only in the QoS constraints, increasing
$p_l^{\mathrm U}$ beyond the minimum level required to satisfy C1 cannot
improve the objective and instead increases the uplink-to-downlink and
uplink multiuser interference. Thus, the uplink-power update
reduces to finding the minimum uplink transmit powers satisfying the QoS
constraints with equality, i.e.,
\begin{equation}
p_l^{\text U,\min}
=
\tfrac{\Gamma}
{
(\bm S \bm h^{\text{UL}}_{l|\text{ext}})^{\text H}
\bm R_l^{-1}
\bm S \bm h^{\text{UL}}_{l|\text{ext}}
},
\quad
\forall l.
\label{eq:min_ul_power}
\end{equation}
In other words, at the optimum, constraint $\text C1$ must be active. Otherwise, if
$
\gamma_l^{\text{UU}}
>
\Gamma
$
while
$
p_l^{\text U}
<
P_{\max,l}^{\text U},
$
the $p_l^{\text U}$ can be further reduced without violating constraint C7, thereby decreasing SI and improving the objective value. Hence, the optimal $p_l^{\text U}$ is 
\begin{equation}
(p_l^{\text U})^\star
=
p_l^{\text U,\min},
\quad
\forall l.
\label{eq:optimal_ul_power}
\end{equation}
Since $\bm R_l$ depends on the powers of all uplink users, we update
$\{p_l^{\mathrm U}\}$ using the following fixed-point iteration:
\begin{equation}
p_l^{\mathrm U(t_{\text{U}}+1)} = p_l^{\mathrm U,\min}, \quad \forall l,
\label{eq:fixed_point_update}
\end{equation}
where $(t_{\text{U}})$ denotes the iteration index for updating $p_l$ and \(p_l^{\mathrm U(t_{\text{U}}+1)}\) is the uplink power at the next iteration. After each update, $\bm R_l$ is recomputed using
\eqref{eq:mmse_combiner}. If
$p_l^{\mathrm U(t_{\text{U}}+1)} > P_{\max,l}^{\mathrm U}$ for any $l$, constraint C1 is infeasible under the current downlink beamformers.
The updates continue until convergence~\cite{yates1995framework}.
\label{sol:WVPs}

\subsection{Block 2: Optimizing Pinching Beamforming and Waveguide Mode Selector} 

For fixed $\{\bm{w}_{k|\mathrm{ext}},\bm{v}_{l|\mathrm{ext}},p_l^{\mathrm U}\}$,
the optimization problem in Block 2 remains highly non-convex due to the logarithmic SINR
objective.
By adopting the FP framework\cite{8310563}, it can be equivalently transformed as 
\begin{align}
  \underset{ \substack{x^{\text{Pin}}_{n,m}, \eta_{n,m},\\s_n,\,\alpha_k,\,\nu_k}}{\text{maximize}} \,\,&f_{\text{FP}}\triangleq\tfrac{1}{\ln 2} \sum_{k=1}^{K}\mu_k 
  \Big[
  \ln(1\hspace{-1mm}+\hspace{-1mm}\alpha_k)\hspace{-1mm}-\hspace{-1mm}\alpha_k-\abs{\nu_k}^2(\abs{\xi^{\text N}_{k}}^2\notag\\[-1mm]&+\sum_{q\neq k}^{K}\abs{\xi^{\text D}_{q,k}}^2+\sigma_{k,\text{Deff}}^2)+
   2\mathfrak{R}\big\{\nu_k^*\sqrt{(1+\alpha_k)}\xi^{\text N}_{k}\big\}
  \Big]\notag \\[-1mm]
  \text{s.t.}\,\,
 & \text{C1}-\text{C6},\text{C8},\text{C9},\label{eq:block2_problem_FP}
\end{align}
where  $\alpha_k \ge 0$ and $\nu_k \in \mathbb{C}$ are auxiliary optimization variables. $
\xi^{\text N}_{k} = \sum_{n=1}^{N} \sum_{m=1}^{M} \sqrt{\rho_0}[\bm{w}_{k|\text{ext}}]_n(1-s_n)\sqrt{\eta_{n,m}} z^{\text{DU}}_{k,n,m}$, $
\xi^{\text D}_{q,k} =  \sum_{n=1}^{N} \sum_{m=1}^{M} \sqrt{\rho_0}[\bm{w}_{q|\text{ext}}]_n(1-s_n)\sqrt{\eta_{n,m}} z^{\text{DU}}_{k,n,m}
$, $
z^{\text{DU}}_{k,n,m}
=
\tfrac{
e^{-j\theta^{\text{DU}}_{k,n,m}}
}{
d^{\text{DU}}_{k,n,m}(x^{\text{Pin}}_{n,m})
}$, and $\theta^{\text{DU}}_{k,n,m}( x^{\text{Pin}}_{n,m})
=
\beta_0 d^{\text{DU}}_{k,n,m}
+
(
\beta_1 x^{\text{Pin}}_{n,m}
-
\sum_{q=1}^{m-1}\Delta \phi_{n,q}
)$.

To tackle \eqref{eq:block2_problem_FP}, we alternately update the auxiliary variables
$\{\alpha_k,\nu_k\}$ and the physical design variables
$\{x_{n,m}^{\mathrm{Pin}},\eta_{n,m},s_n\}$. Specifically, for fixed $\{x_{n,m}^{\mathrm{Pin}},\eta_{n,m},s_n\}$, the optimal auxiliary variables admit the following closed-form expressions
\begin{align}
\alpha_k^{\star}
=&
\tfrac{\abs{\xi^{\text N}_{k}}^2}
{\sum_{q\neq k}^{K}\abs{\xi^{\text D}_{q,k}}^2+\sigma_{k,\text{Deff}}^2},\label{eq:FP_alpha}\\
\nu_k^{\star}
=&
\tfrac{
\sqrt{1+\alpha_k}\xi^{\text N}_{k}
}
{
\abs{\xi^{\text N}_{k}}^2
+
\sum_{q\neq k}^{K}\abs{\xi^{\text D}_{q,k}}^2
+
\sigma_{k,\text{Deff}}^2
}.\label{eq:FP_nu}
\end{align}

With the auxiliary variables fixed, the physical design variables are then
updated by solving
\begin{equation}
\begin{aligned}
  \underset{x^{\text{Pin}}_{n,m}, \eta_{n,m},s_n}{\text{maximize}} \,\,& f_{\text{FP}}\\
  \text{s.t.}\,\,
 & \text{C1}-\text{C6},\text{C8},\text{C9}.
\end{aligned}
\label{eq:block2_problem_FP_1}
\end{equation}
Two variable blocks are updated alternately until convergence.

To handle the coupling between $(s_n,\eta_{n,m})$ and
$x^{\mathrm{Pin}}_{n,m}$ in problem \eqref{eq:block2_problem_FP_1}, we employ
a BCD framework that alternately optimizes each variable block while keeping
the other fixed.
\label{sol:PBs-FP}

\subsubsection{Sub-problem 1} For fixed $\{\bm{w}_{k|\text{ext}}, \bm{v}_{l|\text{ext}}, p_l^{\text U},x_{n,m}^{\text{Pin}}\}$, the resulting subproblem aims to jointly optimize the waveguide mode-selection binary variables $s_n$ and the power radiation variables $\eta_{n,m}$, which is given by
\begin{equation}
\begin{aligned}
  \underset{\eta_{n,m},\,s_n}{\text{maximize}} \,\,& f_{\text{FP}}\\
  \text{s.t.}\,\,
 & \text{C1}, \text{C4}-\text{C6},\text{C8},\text{C9}.
\end{aligned}
\label{eq:block2_problem_FP_SP1}
\end{equation}
To handle the coupling term $s_n\sqrt{\eta_{n,m}}$ in the objective function and constraints C1 and C4, we first define
$
\eta_{n,m}^{\text{s}} \triangleq \sqrt{\eta_{n,m}},  \forall n,m,
$
and introduce an auxiliary slack variable $\widetilde{\eta^{\text{s}}}_{n,m}$ with the following constraint:
\begin{equation}
\text{C10}: \quad
\widetilde{\eta^{\text{s}}}_{n,m}
=
\eta_{n,m}^{\text{s}} s_n,
\quad \forall n,m.
\end{equation}
By applying the big-M reformulation~\cite{hu2021robustSecureSumRateIRS}, the bilinear constraint
C10 can be equivalently transformed as
\begin{equation}
\begin{aligned}
\text{C10a}\hspace{-1mm}:\quad&
\widetilde{\eta^{\text{s}}}_{n,m}
\leq
\eta^{\text{s}}_{n,m},
\forall n,m,
\\\text{C10b}\hspace{-1mm}:\quad&
\widetilde{\eta^{\text{s}}}_{n,m}
\leq
M_{n,m}s_n,
\forall n,m,
\\
\text{C10c}\hspace{-1mm}:\quad&
\widetilde{\eta^{\text{s}}}_{n,m}
\geq
\eta^{\text{s}}_{n,m}
-
M_{n,m}(1-s_n),
\forall n,m,
\\\text{C10d}\hspace{-1mm}:\quad&
\widetilde{\eta^{\text{s}}}_{n,m}
\geq0,
\forall n,m,
\end{aligned}\end{equation}
where $
M_{n,m}=
\sqrt{\text{PL}\big(x^{\text{Pin}}_{n,m}\big)}, \forall
 m$.
%Specifically, when \(s_n=0\), constraint \(\text{C10b}\) together with \(\text{C10d}\) enforces \(\widetilde{\eta^{\text{s}}}_{n,m}=0\). On the other hand, when \(s_n=1\), constraints \(\text{C10a}\) and \(\text{C10c}\) jointly yield \(\widetilde{\eta^{\text{s}}}_{n,m}=\eta^{\text{s}}_{n,m}\). Therefore, constraints \(\text{C10a}\)-\(\text{C10d}\) are equivalent to the original bilinear relation\(\widetilde{\eta^{\text{s}}}_{n,m}=s_n\eta^{\text{s}}_{n,m}\)for binary variable \(s_n\).

Moreover, to deal with the binary constraint  $\text{C8}$, we equivalently transformed it as $\text{C8a}$ and $\text{C8b}$, which are given by
 \begin{equation}
 \text{C8a}\hspace{-1mm}: \sum_{n=1}^{N}(s_n-s_n^2) \leq 0, \text{C8b}\hspace{-1mm}:0\leq s_n\leq 1,\forall n.\\
\end{equation}
%After relaxing $s_n$ to $[0,1]$, constraints C10a-C10d provide a convex McCormick relaxation.
Then, we incorporate
$\text{C8a}$
into the objective function as a penalty term, yielding
\begin{align}
&\max_{\substack{\eta^{\text{s}}_{n,m},\widetilde{\eta^{\text{s}}}_{n,m},s_n}}
\quad  f_{\text{FP,SP1}} -\rho_{\text{penalty}} \sum_{n=1}^N (s_n-s_n^2) \label{eq:FPSUB1}\\
\mathrm{s.t.}\quad& \text{C1}\hspace{-1mm}:  
\Gamma
\big(f^{\text{SI}}_{l}
+
(\sigma^{\text{BS}})^2\sum_{n=1}^{N}\big|[\bm{v}_{l|\text{ext}}^{\text{H}}]_n\big|^2 s_n^2
\notag\\&\hspace{10mm}+
\sum_{q\neq l}^{L}
p_q^{\text U}\big|
\sum_{n=1}^{N}
\rho_0^{\tfrac{1}{2}}
([\bm{v}_{l|\text{ext}}]_n)^*
\widetilde{\eta^{\text{s}}}_{n,1}
z^{\text{UU}}_{q,n}
\big|^2
\big)
\notag\\& \hspace{10mm}\leq
p_l^{\text U}
\big|
\sum_{n=1}^{N}
\rho_0^{\tfrac{1}{2}}
([\bm{v}_{l|\text{ext}}]_n)^*
\widetilde{\eta^{\text{s}}}_{n,1}
z^{\text{UU}}_{l,n}
\big|^2, \forall l,\notag\\
&\text{C4a}\hspace{-1mm}:
0\le (\eta^{\text{s}}_{n,m})^2\le\text{PL}\big(x^{\text{Pin}}_{n,m}\big)(1-s_n),
\notag\\&
\text{C4b}\hspace{-1mm}:
(\eta^{\text{s}}_{n,m})^2
\le
\text{PL}\big(x^{\text{Pin}}_{n,m}\big)\hspace{-1mm}
-\hspace{-1mm}
\sum_{i=1}^{m-1}
(\eta^{\text{s}}_{n,i})^2
\text{PL}\big(
x^{\text{Pin}}_{n,m}\hspace{-1mm}-x^{\text{Pin}}_{n,i}
\big),\notag\\
& \text{C5}\hspace{-1mm}:0\leq (\eta^{\text{s}}_{n,1})^2\leq (1 - s_n)\text{PL}\big(x^{\text{Pin}}_{n,1}\big)+s_n\eta^\text{UL}_{\max}\text{PL}(x^{\text{Pin}}_{n,1}),\notag\\[-0mm]&\text{C6},\text{C8b},\text{C9},\text{C10a}-\text{C10d},\notag
\end{align}
where 
\begin{align}f^{\text{SI}}_{l}\triangleq&\rho_{\text{ANA}}
\sum_{k=1}^{K}
\big|
\sqrt{\rho_0}
\sum_{n=1}^{N}\sum_{i=1,i\neq n}^{N}
\sum_{m=1}^{M}
([\bm{v}_{l|\text{ext}}]_i)^*
[\bm{w}_{k|\text{ext}}]_n\notag\\&\times
\left(
\widetilde{\eta^{\text{s}}}_{i,1}\eta^{\text{s}}_{n,m}
-
\widetilde{\eta^{\text{s}}}_{i,1}
\widetilde{\eta^{\text{s}}}_{n,m}
\right)
z^{\text{SI}}_{i,n,m}
\big|^2,  \\
f_{\text{FP,SP1}} \triangleq&
\tfrac{1}{\ln 2}
\sum_{k=1}^{K}\mu_k
\big[
\ln(1+\alpha_k)-\alpha_k+
2\mathfrak{R}
\big\{
\nu_k^*\sqrt{1+\alpha_k}\notag\\&
\hspace{-12mm}\times
\sum_{n=1}^{N}\sum_{m=1}^{M}
\rho_0^{\tfrac{1}{2}}
z^{\text{DU}}_{k,n,m}
[\bm{w}_{k|\text{ext}}]_n
(\eta^{\text{s}}_{n,m}-\widetilde{\eta^{\text{s}}}_{n,m})\big\}
\hspace{-1mm}-|\nu_k|^2
\big(\sigma_{k,\text{Deff}}^2
\notag\\&
\hspace{-8mm}+
\hspace{-1mm}\sum_{q=1}^{K}
\hspace{-0mm}\big|
\hspace{-0mm}\sum_{n=1}^{N}\hspace{-0mm}\sum_{m=1}^{M}
\sqrt{\rho_0}
z^{\text{DU}}_{k,n,m}
[\bm{w}_{q|\text{ext}}]_n
(\eta^{\text{s}}_{n,m}\hspace{-0mm}-\hspace{-0mm}\widetilde{\eta^{\text{s}}}_{n,m})\big|^2
\big)
\big],\notag
\end{align}
$z^{\text{UU}}_{l,n}
=
\tfrac{
e^{-j\theta^{\text{UU}}_{l,n}}
}{
d^{\text{UU}}_{l,n}(x^{\text{Pin}}_{n,1})
},
 \theta^{\text{UU}}_{l,n}( x^{\text{Pin}}_{n,1})
=
\beta_1 x^{\text{Pin}}_{n,1}
+
\beta_0 d^{\text{UU}}_{l,n}, z^{\text{SI}}_{i,n,m}
=
\tfrac{
e^{-j\theta^{\text{SI}}_{i,n,m}}
}{
d^{\text{SI}}_{i,n,m}(x^{\text{Pin}}_{n,m})
}$, and 
$\theta^{\text{SI}}_{i,n,m}( x^{\text{Pin}}_{n,m})
=\beta_1x^{\text{Pin}}_{i,1}
+
\beta_0 d^{\text{SI}}_{i,n,m}
+
\beta_1 x^{\text{Pin}}_{n,m}
-
\sum_{q=1}^{m-1}\Delta \phi_{n,q}), \forall i\neq n.$
Here, $\rho_{\text{penalty}}\gg 1$ is a sufficiently large penalty factor.
It can be shown that the penalized problem is equivalent
to the original problem in \eqref{eq:block2_problem_FP_SP1} when $\rho_{\text{penalty}}$ is sufficiently large \cite{9449661}.

 Since the non-convexity of constraint $\text{C1}$ in \eqref{eq:FPSUB1} came from the bilinear coupling term 
$
\widetilde{\eta^\text{s}}_{i,1} \eta^\text{s}_{n,m}
-
\widetilde{\eta^\text{s}}_{i,1}\widetilde{\eta^\text{s}}_{n,m},
$ appeared in the SI term $f^{\text{SI}}_{l}$, we introduce an auxiliary variable $\eta^{\text{SI}}_{i,n,m}$ and construct a conservative upper bound of the SI power via the triangle inequality. Specifically, constraint C1 is replaced by
  \begin{align}  
\text{C1a}\hspace{-1mm}: & \,\Gamma
\Big(\sum_{q\neq l}^{L}p^{\text{U}}_{q}\big|\sum_{n=1}^{N}\rho_0^{\tfrac{1}{2}}([\bm{v}_{l|\text{ext}}]_n)^{*}\widetilde{\eta^\text{s}}_{n,1} z^{\text{UU}}_{q,n}\big|^2+
\widehat{f}^{\text{SI}}_{l}+(\sigma^{\text{BS}})^2\notag\\&\hspace{-6mm}\times\sum_{n=1}^{N}|[\bm{v}_{l|\text{ext}}^{\text{H}}]_n|^2 s_n^2
\big)
\Big)\leq  p^{\text{U}}_{l}\big|\sum_{n=1}^{N}\rho_0^{\tfrac{1}{2}}([\bm{v}_{l|\text{ext}}]_n)^{*}\widetilde{\eta^\text{s}}_{n,1}  z^{\text{UU}}_{l,n}\big|^2 , \notag\\[-0mm]
\text{C1b}\hspace{-1mm}:&\, \eta^{\text{SI}}_{i,n,m} \geq \
\widetilde{\eta^\text{s}}_{i,1} \eta^\text{s}_{n,m}- \widetilde{\eta^\text{s}}_{i,1}  \widetilde{\eta^\text{s}}_{n,m}, \notag\\[-0mm]
\text{C1c}\hspace{-1mm}:&\,\eta^{\text{SI}}_{i,n,m} \geq 0,\end{align}
where $f_l^{\text{SI}}
\le
\widehat f_l^{\text{SI}}
\triangleq\rho_{\text{ANA}}
\sum_{k=1}^{K}\big(
\sqrt{\rho_0}
\sum_{n=1}^{N}\sum_{\substack{i=1 \\i\neq n}}^{N}
\sum_{m=1}^{M}$\\$
\abs{([\bm{v}_{l|\text{ext}}]_i)^*
[\bm{w}_{k|\text{ext}}]_n z^{\text{SI}}_{i,n,m}}
\eta^{\text{SI}}_{i,n,m}
\big)^2.$
Note that, although constraints C1a–C1c may exclude some feasible solutions of the original problem \eqref{eq:FPSUB1}, any solution satisfying them is guaranteed to fulfill the required UL-QoS constraints.

Then, we apply first-order Taylor approximations to constraints C1a and C1b and replace $s_n^2$ in the objective function with its affine lower bound. At the $(t_{\text{MM}})$-th iteration, the resulting convex surrogate constraints,
denoted by $\widetilde{\mathrm{C1a}}$ and $\widetilde{\mathrm{C1b}}$, are given
by
\begin{align}
\widetilde{\text{C1a}}\hspace{-1mm}: & \Gamma
\big(
\sum_{q\neq l}^{L}
p^{\text{U}}_{q}
\big|
\sum_{n=1}^{N}
\rho_0^{\tfrac{1}{2}}
([\bm{v}_{l|\text{ext}}]_n)^{*}
\widetilde{\eta^\text{s}}_{n,1}
z^{\text{UU}}_{q,n}
\big|^2+\widehat f_l^{\text{SI}}\notag\\&
+
(\sigma^{\text{BS}})^2\sum_{n=1}^{N}|[\bm{v}_{l|\text{ext}}^{\text{H}}]_n|^2 s^2_n
\big)
\leq
f^{\text{MM}}_\text{C1a},
\quad \forall l,
\\
\widetilde{\text{C1b}}\hspace{-1mm}:&
\eta^{\text{SI}}_{i,n,m}
\geq f^{\text{MM}}_\text{C1b}
\quad \forall i,n,m, \text{ where} \notag
\end{align}
\begin{align}
&f^{\text{MM}}_{\text{C1a}} \hspace{-1mm}\triangleq\hspace{-1mm}
p^{\text{U}}_{l}\rho_0\Big(
|
\hspace{-1mm}\sum_{n=1}^{N}
([\bm{v}_{l|\text{ext}}]_n)^*
\widetilde{\eta^{\text{s}}}_{n,1}^{(t_{\text{MM}})}
z^{\text{UU}}_{l,n}
|^2\hspace{-1mm}
+\hspace{-1mm}
2
\Re
\{
(
\hspace{-1mm}\sum_{n=1}^{N}([\bm{v}_{l|\text{ext}}]_n)^*
 \notag\\&
\times\hspace{-1mm}
\widetilde{\eta^{\text{s}}}_{n,1}^{(t_{\text{MM}})}
z^{\text{UU}}_{l,n}
)^*
(
\sum_{n=1}^{N}
([\bm{v}_{l|\text{ext}}]_n)^*
\big(
\widetilde{\eta^\text{s}}_{n,1}
-
\widetilde{\eta^{\text{s}}}_{n,1}^{(t_{\text{MM}})}
\big)
z^{\text{UU}}_{l,n}
)
\}\Big)\\
&f^{\text{MM}}_\text{C1b} \triangleq
\tfrac{1}{4}
(\widetilde{\eta^\text{s}}_{i,1}
+
\eta^\text{s}_{n,m})^2
+\tfrac{1}{4}
(\widetilde{\eta^\text{s}}_{i,1}
-
\widetilde{\eta}^\text{s}_{n,m})^2
+\tfrac{1}{4}
(\widetilde{\eta^{\text{s}}}_{i,1}^{(t_{\text{MM}})}\notag
\\& -
\eta^{\text{s}(t_{\text{MM}})}_{n,m})^2
-\tfrac{1}{2}
(\widetilde{\eta^{\text{s}}}_{i,1}^{(t_{\text{MM}})}
-
\eta^{\text{s}(t_{\text{MM}})}_{n,m})
(
\widetilde{\eta^\text{s}}_{i,1}
-\eta^\text{s}_{n,m})
+\tfrac{1}{4}(\widetilde{\eta^{\text{s}}}_{i,1}^{(t_{\text{MM}})}\notag\\
&%\hspace{9mm}
+
\widetilde{\eta^{\text{s}}}_{n,m}^{(t_{\text{MM}})}
)^2
-\tfrac{1}{2}(
\widetilde{\eta^{\text{s}}}_{i,1}^{(t_{\text{MM}})}+
\widetilde{\eta^\text{s}}_{n,m}^{(t_{\text{MM}})})(\widetilde{\eta^{\text{s}}}_{i,1}+\widetilde{\eta^\text{s}}_{n,m})\notag
\end{align}
 
Thus, a lower bound of  \eqref{eq:FPSUB1} is obtained by solving the following problem in the $(t_{\text{MM}})$-th MM iteration:
\begin{align}
\max_{\substack{\eta^{\text{s}}_{n,m},\,\widetilde{\eta^{\text{s}}}_{n,m},\\\eta^{\text{SI}}_{i,n,m},\,s_n}}
\quad & f_{\text{FP,SP1}} \hspace{-1mm}-\hspace{-1mm}\rho_{\text{penalty}} \sum_{n=1}^N (s_n\hspace{-1mm}- \hspace{-1mm}(2s_n^{(t_{\text{MM}})} s_n
\hspace{-1mm}-\hspace{-1mm}
(s_n^{(t_{\text{MM}})})^2))\notag \\
s.t.\quad& \widetilde{\text{C1a}}, \widetilde{\text{C1b}},\text{C1c},
\text{C4a}, \text{C4b}, \text{C5},\label{eq:FP-SP1}\\&\text{C6},\text{C8b},\text{C9},\text{C10a}-\text{C10d}.\notag
\end{align}
Now \eqref{eq:FP-SP1} is a convex problem and can be solved by adopting a standard numerical solver for convex programming. \label{sol:PBs-ETAS}
\subsubsection{Subproblem 2} 
With fixed variables $\{w_{k|\text{ext}}, v_{l|\text{ext}}, p_l^U, s_n, \eta_{n,m}\}$, the optimization problem in \eqref{eq:block2_problem_FP_1} can be equivalently rewritten as
\begin{align}
  \underset{x^{\text{Pin}}_{n,m}}{\text{maximize}} \,\,&f^{\text{ADMM}}_{\text{obj}}\label{eq:block2a_problem_X}
\\
  \text{s.t.}\,\,
  & \text{C1}\hspace{-1mm}:f^{\text{ADMM}}_{\text{C1},l}\leq 0,
\forall l,\text{C2}-\text{C5},\notag
\end{align}
where \vspace{-2mm}\begin{align}f^{\text{ADMM}}_{\text{obj}}&\triangleq
  \tfrac{1}{\ln 2} \sum_{k=1}^{K}\mu_k 
  \Big[
  \ln(1+\alpha_k)-\alpha_k+
  2\mathfrak{R}\big\{\nu_k^*\sqrt{(1+\alpha_k)} \notag\\[-2mm] &\hspace{-11mm}\times\sum_{n=1}^{N} \sum_{m=1}^{M} c^{\text{DL}}_{k,n}\sqrt{\eta_{n,m}}z^{\text{DU}}_{k,n,m}\big\}-
  |\nu_k|^2
  \Big(
\big|
  \sum_{n=1}^{N} \sum_{m=1}^{M} 
  c^{\text{DL}}_{k,n}\sqrt{\eta_{n,m}}\notag\\[-2mm] &\hspace{-9mm} \times z^{\text{DU}}_{k,n,m}
  \big|^2 +\sigma_{k,\text{Deff}}^2+
\sum_{q\neq k}^{K}
\big|
\sum_{n=1}^{N} \sum_{m=1}^{M}
c^{\text{DL}}_{q,n}\sqrt{\eta_{n,m}}z^{\text{DU}}_{k,n,m}
\big|^2
\Big)
  \Big], \notag\\[-0mm]
 f^{\text{ADMM}}_{\text{C1},l}&\triangleq\hspace{-1mm}\Gamma
\Big(
\rho_{\text{ANA}}\hspace{-1mm}
\sum_{k=1}^{K}
\Big|
\sum_{i\neq n}^{N}
\sum_{n=1}^{N}
\sum_{m=1}^{M}
c^{\text{SI}}_{l,k,i,n}
\sqrt{\eta_{i,1}\eta_{n,m}}
z^{\text{SI}}_{i,n,m}
\Big|^2\hspace{-2mm} \notag \\[-2mm]&\hspace{-9mm}+\sum_{q\neq l}^{L}
p^{\text{U}}_{q}
\Big|
\sum_{n=1}^{N}
c^{\text{UU}}_{l,n}
\sqrt{\eta_{n,1}}
z^{\text{UU}}_{q,n}
\Big|^2
+
(\sigma^{\text{BS}})^2\sum_{n=1}^{N}|[\bm{v}_{l|\text{ext}}^{\text{H}}]_n|^2 s_n^2
\Big)\notag\\[-2mm]
&-p^{\text{U}}_{l}
\Big|
\sum_{n=1}^{N}
c^{\text{UU}}_{l,n}
\sqrt{\eta_{n,1}}
z^{\text{UU}}_{l,n}
\Big|^2, \notag\\[-9mm]\notag
\end{align}
$c^{\text{DL}}_{k,n} = [\overline{\bm{S}}\bm{w}_{k|\text{ext}}]_n\rho_0^{\tfrac{1}{2}}$, $c^{\text{UU}}_{l,n}=[\bm{v}_{l|\text{ext}}^{\text{H}}\bm{S}]_n \rho_0^{\tfrac{1}{2}}$, and $c^{\text{SI}}_{l,k,i,n}=\sqrt{\rho_0}[\bm{v}_{l|\text{ext}}]_i^{*}
\,s_i(1-s_n)\,
[\bm{w}_{k|\text{ext}}]_n$.

For ADMM-based variable splitting, we introduce local channel copies 
$
\widetilde{\mathcal Z}\triangleq
\{\widetilde{z}^{\mathrm{DU}}_{k,n,m},
\widetilde{z}^{\mathrm{UU}}_{l,n},
\widetilde{z}^{\mathrm{SI}}_{i,n,m}\}$, phase copies $
\bm{\Phi}\triangleq
\{\phi^{\mathrm{DU}}_{k,n,m},
\phi^{\mathrm{UU}}_{l,n},
\phi^{\mathrm{SI}}_{i,n,m}\},
$ and consensus constraints C11 and C12, yielding the equivalent reformulation:
\begin{align}
\label{eq:admm_reformulation}
&\max_{\bm{\Phi},{x}^{\text{Pin}}_{n,m},\widetilde{\mathcal Z}}
\quad
f^{\text{ADMM}}_{\text{obj}}(\widetilde{\mathcal Z})
\\
&\text{s.t.}\hspace{2mm}
\text{C1}\hspace{-1mm}: f^{\text{ADMM}}_{\text{C1},l}(\widetilde{\mathcal Z})\leq 0, 
\text{C2}-\text{C5},\notag
\\&
\hspace{6mm}
\text{C11}\hspace{-1mm}:
\chi^{\text{DU}}_{k,n,m} = 0,
\chi^{\text{UU}}_{l,n} = 0, 
\chi^{\text{SI}}_{i,n,m} =0, \forall k,l,i,n,m,\notag\\&\hspace{6mm}\text{C12}\hspace{-1mm}:
\varphi^{\text{DU}}_{k,n,m}
=
0,
\varphi^{\text{UU}}_{l,n}=
0,
\varphi^{\text{SI}}_{i,n,m}
=0, \forall i,n,m,\notag
\end{align}
where $\chi^{\text{DU}}_{k,n,m} = \widetilde{z^{\text{DU}}}_{k,n,m}d^{\text{DU}}_{k,n,m}({x}^{\text{Pin}}_{n,m}) - 
e^{-j\Phi^{\text{DU}}_{k,n,m}}$, $\chi^{\text{UU}}_{l,n}  = \widetilde{z^{\text{UU}}}_{l,n}d^{\text{UU}}_{l,n}({x}^{\text{Pin}}_{n,1}) -e^{-j\Phi^{\text{UU}}_{l,n}}$, $\chi^{\text{SI}}_{i,n,m}=\widetilde{z^{\text{SI}}}_{i,n,m}d^{\text{SI}}_{i,n,m}({x}^{\text{Pin}}_{n,m})-
e^{
-j\Phi^{\text{SI}}_{i,n,m} 
}$, $\varphi^{\text{DU}}_{k,n,m}=\phi^{\text{DU}}_{k,n,m}
-
\theta^{\text{DU}}_{k,n,m}(x^{\text{Pin}}_{n,m})$, $\varphi^{\text{UU}}_{l,n}
=\phi^{\text{UU}}_{l,n}-
\theta^{\text{UU}}_{l,n}(x^{\text{Pin}}_{n,1})$, and $\varphi^{\text{SI}}_{i,n,m}
=\phi^{\text{SI}}_{i,n,m}-\theta^{\text{SI}}_{i,n,m}( x^{\text{Pin}}_{n,m}), \forall i,n,m$.
By introducing the scaled dual variables
$\bm U^{\text{P1}}\triangleq
\{
u^{\text{DU,P1}}_{k,n,m},
u^{\text{UU,P1}}_{l,n},
u^{\text{SI,P1}}_{i,n,m}
\}$ and
$\bm U^{\text{P2}}\triangleq
\{
u^{\text{DU,P2}}_{k,n,m},
u^{\text{UU,P2}}_{l,n},
u^{\text{SI,P2}}_{i,n,m}
\}$
associated with constraints C11 and C12, respectively, and the corresponding ADMM penalty parameters
$\rho_1>0$ and
$\rho_2>0$,
the scaled augmented Lagrangian associated with
\eqref{eq:admm_reformulation}
is given by
\begin{align}
\mathcal L_\rho
(
\bm{\Phi},
{x}^{\text{Pin}}_{n,m},
\widetilde{\mathcal Z},
\bm U^{\text{P1}}\hspace{-1mm},\bm U^{\text{P2}}
)\hspace{-1mm}
=&\hspace{-1mm}
-\hspace{-1mm}f^{\text{ADMM}}_{\text{obj}}(\widetilde{\mathcal Z})
\hspace{-1mm}+\hspace{-1mm}f_{\text P2}
(
\bm{\Phi},{x}^{\text{Pin}}_{n,m},
\bm U^{\text{P2}}
)
\notag\\&+
f_{\text P1}
(
\widetilde{\mathcal Z},\bm{\Phi},{x}^{\text{Pin}}_{n,m},
\bm U^{\text{P1}}
),\label{eq:scaled_augmented_lagrangian}
\end{align}
subject to constraints
$
\text{C1-C5}
$, 
where
\begin{align}
f_{\text P1}
\triangleq
&
\tfrac{\rho_1}{2}\Big(\sum_{k=1}^{K}
\sum_{n=1}^{N}
\sum_{m=1}^{M}
|
 \chi^{\text{DU}}_{k,n,m}
\hspace{-0mm}+\hspace{-0mm}
u^{\text{DU,P1}}_{k,n,m}
|^2
\hspace{-0mm}+\sum_{l=1}^{L}
\sum_{n=1}^{N}
|
 \chi^{\text{UU}}_{l,n}
\hspace{-0mm}\notag\\&\hspace{-6mm}
+\hspace{-0mm}
u^{\text{UU,P1}}_{l,n}
|^2\hspace{-0mm}+
\sum_{\substack{i=1\\i\neq n}}^{N}
\sum_{n=1}^{N}
\sum_{m=1}^{M}
|
 \chi^{\text{SI}}_{i,n,m}
+
u^{\text{SI,P1}}_{i,n,m}
|^2\Big),\label{eq:admm-fp1}
\\
f_{\text P2}
\triangleq
&
\tfrac{\rho_2}{2}\Big(
\sum_{k=1}^{K}
\sum_{n=1}^{N}
\sum_{m=1}^{M}
|
\varphi^{\text{DU}}_{k,n,m}
+
u^{\text{DU,P2}}_{k,n,m}
|^2
+\sum_{l=1}^{L}
\sum_{n=1}^{N}
|
\varphi^{\text{UU}}_{l,n}\notag\\&
\hspace{-6mm}
+
u^{\text{UU,P2}}_{l,n}
|^2+
\sum_{\substack{i=1\\i\neq n}}^{N}
\sum_{n=1}^{N}
\sum_{m=1}^{M}
|
\varphi^{\text{SI}}_{i,n,m}
+
u^{\text{SI,P2}}_{i,n,m}
|^2\Big).\label{eq:admm-fp2}
\end{align}
Then, we iteratively update variables $\bm{\Phi},{x}^{\text{Pin}}_{n,m},\widetilde{\mathcal Z}$:

\noindent\textbf{ADMM step 1}:
\begin{align}
\widetilde{\mathcal Z}^{t_{\text{A}}+1}
=
\arg\min
&
\mathcal L_\rho (
\bm{\Phi}^{(t_{\text{A}})},
{x}^{\text{Pin}(t_{\text{A}})}_{n,m},
\widetilde{\mathcal Z},
\bm U^{\text{P1}(t_{\text{A}})},\bm U^{\text{P2}(t_{\text{A}})}
)\notag
\\
\text{s.t.}
&\quad
\text{C1}.\label{eq:ADMM_block1_update}
\end{align}

 \noindent\textbf{ADMM step 2}: 
\begin{align}
\hspace{-1mm}\bm\Phi^{t_{\text{A}}+1}
\hspace{-1mm}=\hspace{-1mm}
\arg\min
\mathcal L_\rho(
\bm{\Phi}\hspace{-1mm},
{x}^{\text{Pin}(t_{\text{A}})}_{n,m}\hspace{-1mm},
\widetilde{\mathcal Z}^{(t_{\text{A}}+1)}\hspace{-1mm},
\bm U^{\text{P1}(t_{\text{A}})}\hspace{-1mm},\bm U^{\text{P2}(t_{\text{A}})}
).\hspace{-1mm}\label{eq:ADMM_block2_update}
\end{align}

 \noindent  \textbf{ADMM step 3}: 
\begin{align}
{x}^{\text{Pin}(t_{\text{A}}+1)}_{n,m}
=
\arg\min
&
\mathcal L_\rho(
\bm{\Phi}^{(t_{\text{A}}+1)},
{x}^{\text{Pin}}_{n,m},
\widetilde{\mathcal Z}^{(t_{\text{A}}+1)},\notag\\&
\bm U^{\text{P1}(t_{\text{A}})},\bm U^{\text{P2}(t_{\text{A}})}
)\notag \\
\text{s.t.}\quad
\text{C2}-\text{C5}.\label{eq:ADMM_block3_update}
\end{align}
 \noindent \textbf{ADMM-step 4}: 
\begin{align}
u^{\text{DU,P1}(t_{\text{A}}+1)}_{k,n,m}
&=
u^{\text{DU,P1}(t_{\text{A}})}_{k,n,m}
+
\chi^{{\text{DU}}^{(t_{\text{A}}+1)}}_{k,n,m},
\label{eq:dual_p2}\\
u^{\text{UU,P1}(t_{\text{A}}+1)}_{l,n}
&=
u^{\text{UU,P1}(t_{\text{A}})}_{l,n}
+
\chi^{{\text{UU}}^{(t_{\text{A}}+1)}}_{l,n},
\nonumber\\
u^{\text{SI,P1}(t_{\text{A}}+1)}_{i,n,m}
&=
u^{\text{SI,P1}(t_{\text{A}})}_{i,n,m}
+
\chi^{{\text{SI}}^{(t_{\text{A}}+1)}}_{i,n,m}
,\notag\\
u^{\text{DU,P2}(t_{\text{A}}+1)}_{k,n,m}
&=
u^{\text{DU,P2}(t_{\text{A}})}_{k,n,m}
+
\varphi^{\text{DU}(t_{\text{A}}+1)}_{k,n,m},
\nonumber\\
u^{\text{UU,P2}(t_{\text{A}}+1)}_{l,n}
&=
u^{\text{UU,P2}(t_{\text{A}})}_{l,n}
+
\varphi^{\text{UU}(t_{\text{A}}+1)}_{l,n},
\nonumber\\
u^{\text{SI,P2}(t_{\text{A}}+1)}_{i,n,m}
&=
u^{\text{SI,P2}(t_{\text{A}})}_{i,n,m}
+
\varphi^{\text{SI}(t_{\text{A}}+1)}_{i,n,m}.\notag%\\[-11mm]\notag
\end{align}
\paragraph{Solution for ADMM Step 1}
To tackle the non-convex constraint C1 in \eqref{eq:ADMM_block1_update}, we first establish the lower bounds for $|
\sum_{n=1}^{N}
c^{\text{UU}}_{l,n}
\sqrt{\eta_{n,1}}
\widetilde z^{\text{UU}}_{l,n}
|^2$ at
$ \widetilde z^{\text{UU}(t_{\text{A1}})}_{l,n}
$:
\begin{align}
\Big|
\sum_{n=1}^{N}
c^{\text{UU}}_{l,n}
\sqrt{\eta_{n,1}}
\widetilde z^{\text{UU}}_{l,n}
\big|^2 \geq f^{\text{ADMM}}_{\text{MM,C1}},
\end{align}
where $f^{\text{ADMM}}_{\text{MM,C1}}\triangleq
\big(
-
\big|
\sum_{n=1}^{N}
c^{\text{UU}}_{l,n}
\sqrt{\eta_{n,1}}
\widetilde z^{\text{UU}(t_{\text{A1}})}_{l,n}
\big|^2\\+2\Re
\big\{
\big(
\sum_{n=1}^{N}
c^{\text{UU}}_{l,n}
\sqrt{\eta_{n,1}}
\widetilde z^{\text{UU}(t_{\text{A1}})}_{l,n}
\big)^*
\big(
\sum_{n=1}^{N}
c^{\text{UU}}_{l,n}
\sqrt{\eta_{n,1}}
\widetilde z^{\text{UU}}_{l,n}
\big)
\big\}
\big).$
By replacing the non-convex terms in constraints C1 with their first-order lower bounds obtained at the current iterate, a convex inner approximation of \eqref{eq:ADMM_block1_update}  is obtained as
\begin{align}
\min_{\widetilde{\mathcal Z}}
\hspace{2mm}
&
-\hspace{-1mm}f^{\text{ADMM}}_{\text{obj}}(\widetilde{\mathcal Z})
\hspace{-1mm}+\hspace{-1mm}
f_{\text P1}
(
\widetilde{\mathcal Z},\bm{\Phi}^{(t_{\text{A1}})}, {x}^{\text{Pin}(t_{\text{A1}})}_{n,m},
\bm U^{\text{P1}(t_{\text{A1}})},\bm U^{\text{P2}(t_{\text{A1}})}
)
\notag \\
\text{s.t.}
\quad
&
\overline{\text{C1}}:
 p_l^{\text U}f^{\text{ADMM}}_{\text{MM,C1}}
\ge
\Gamma
\big(
\sum_{q\neq l}^{L}
p_q^{\text U}
|
\sum_{n=1}^{N}
c^{\text{UU}}_{l,n}
\sqrt{\eta_{n,1}}
\widetilde z^{\text{UU}}_{q,n}
|^2
\notag \\+
& 
\rho_{\text{ANA}}\sum_{k=1}^{K}
|
\sum_{i\neq n}^{N}
\sum_{n=1}^{N}
\sum_{m=1}^{M}
c^{\text{SI}}_{l,k,i,n}
\sqrt{\eta_{i,1}\eta_{n,m}}
\widetilde z^{\text{SI}}_{i,n,m}
|^2
\notag\\&+
(\sigma^{\text{BS}})^2\sum_{n=1}^{N}|[\bm{v}_{l|\text{ext}}^{\text{H}}]_n|^2 s_n^2
\big),
\forall l.\label{eq:ADMM_block1_update_sca}
\end{align}
\paragraph{Solution for ADMM Step 2}
Since all phase variables are separable in \eqref{eq:ADMM_block2_update}, \eqref{eq:ADMM_block2_update} can be decomposed into independent scalar subproblems. For the variable
\(
\phi^{\text{DU}}_{k,n,m}
\),
the corresponding optimization problem is expressed as
\begin{equation}
\begin{aligned}
\min_{\phi^{\text{DU}}_{k,n,m}}
\tfrac{\rho_1}{2}
|
\varrho^{\text{DU}}_{k,n,m}
\hspace{-1mm}-\hspace{-1mm}
e^{-j\phi^{\text{DU}}_{k,n,m}}
|^2\hspace{-1mm}
+\hspace{-1mm}
\tfrac{\rho_2}{2}
(
\phi^{\text{DU}}_{k,n,m}
\hspace{-1mm}-\hspace{-1mm}
\psi^{\text{DU}}_{k,n,m}
)^2\hspace{-1mm},
\end{aligned}
\label{eq:scalar_phi_problem}
\end{equation}
where
$
\varrho^{\text{DU}}_{k,n,m}
\triangleq
\widetilde z^{\text{DU}}_{k,n,m}
d^{\text{DU}}_{k,n,m}
\left(
{x}^{\text{Pin}}_{n,m}
\right)
+
u^{\text{DU,P1}}_{k,n,m},
$
and
$
\psi^{\text{DU}}_{k,n,m}
\triangleq
\theta^{\text{DU}}_{k,n,m}
-
u^{\text{DU,P2}}_{k,n,m}.
$
The objective function in \eqref{eq:scalar_phi_problem}
can be equivalently rewritten as
\begin{align}
f^{\text{DU,P2}}_{k,n,m}
(
\phi^{\text{DU}}_{k,n,m}
)
=
&
-
\rho_1
\left|
\varrho^{\text{DU}}_{k,n,m}
\right|
\cos
\left(
\phi^{\text{DU}}_{k,n,m}
+
\angle
\varrho^{\text{DU}}_{k,n,m}
\right)
\notag\\&\hspace{-2mm}+
\tfrac{\rho_2}{2}
\left(
\phi^{\text{DU}}_{k,n,m}-
\psi^{\text{DU}}_{k,n,m}
\right)^2
+
C^{\text{DU}}_{k,n,m},\label{eq:gphi_cosine_form}
\end{align}
where
$
C^{\text{DU}}_{k,n,m}
=
\tfrac{\rho_1}{2}
(
|
\varrho^{\text{DU}}_{k,n,m}
|^2
+
1
).
$
The scalar problem in \eqref{eq:scalar_phi_problem} is smooth but does
not admit a closed-form solution because its stationary equation
contains both linear and trigonometric terms. It can therefore be solved
using Newton's method~\cite{boyd2004convex}:
\begin{equation}\hspace{-2mm}
\begin{aligned}
\phi^{\text{DU}(t_{\text{A2}}+1)}_{k,n,m}\hspace{-1mm}
=
\phi^{\text{DU}(t_{\text{A2}})}_{k,n,m}\hspace{-1mm}
-\hspace{-1mm}
{
\tfrac{
\partial
f^{\text{DU,P2}}_{k,n,m}
(
\phi^{\text{DU}(t_{\text{A2}})}_{k,n,m}
)
}{
\partial
\phi^{\text{DU}}_{k,n,m}
}
}/{
\tfrac{
\partial^2
f^{\text{DU,P2}}_{k,n,m}
(
\phi^{\text{DU}(t_{\text{A2}})}_{k,n,m}
)
}{
\partial
(
\phi^{\text{DU}}_{k,n,m}
)^2
}
}.\hspace{-1mm}
\end{aligned}
\label{eq:newton_phi_update}
\end{equation}
The derivatives in \eqref{eq:newton_phi_update} follow directly from
\eqref{eq:gphi_cosine_form} and are omitted for the sake of brevity. When
$\rho_2>\rho_1|\varrho^{\mathrm{DU}}_{k,n,m}|$, the objective in
\eqref{eq:scalar_phi_problem} is strictly convex, yielding a unique optimal
solution. The same element-wise update applies to
$\phi^{\mathrm{UU}}_{l,n}$ and $\phi^{\mathrm{SI}}_{i,n,m}$, completing the
update of $\bm{\Phi}$.
%The above element-wise update is performed until convergence of the inner Newton iterations, and the resulting phase vector is then used in the outer ADMM loop.

\paragraph{Solution for ADMM Step 3}
The non-convexity in \eqref{eq:ADMM_block3_update} arises from both $f_{\text P1}$ and $f_{\text P2}$.
We first handle the nonlinear square-root distance functions in $f_{\text P2}$, i.e., $d^{\text{DU}}_{k,n,m},d^{\text{UU}}_{l,n}, d^{\text{SI}}_{i,n,m}$ in \eqref{eq:admm-fp2}. 
To obtain a tractable reformulation, we introduce the auxiliary variables
$\epsilon^{\text{DU}}_{k,n,m}$,
$\epsilon^{\text{UU}}_{l,n}$,
and
$\epsilon^{\text{SI}}_{i,n,m}$,
and equivalently reformulate the absolute-value terms in $f_{\text P2}$ as
\begin{align}
\text{C13a}\hspace{-1mm}:&
-
\epsilon^{\text{DU}}_{k,n,m}
\le
\phi^{\text{DU}}_{k,n,m} \hspace{-1mm}+ \hspace{-1mm}\widetilde{u^{\text{DU,P2}}_{k,n,m}}
-
\beta_0
d^{\text{DU}}_{k,n,m}
-
\beta_1
x^{\text{Pin}}_{n,m},\notag\\
\text{C13b}\hspace{-1mm}:&
\phi^{\text{DU}}_{k,n,m} + \widetilde{u^{\text{DU,P2}}_{k,n,m}}
-
\beta_0
d^{\text{DU}}_{k,n,m}
-
\beta_1
x^{\text{Pin}}_{n,m}
\le
\epsilon^{\text{DU}}_{k,n,m},\notag\\
\text{C14a}\hspace{-1mm}:&
-
\epsilon^{\text{UU}}_{l,n}
\le 
\phi^{\text{UU}}_{l,n} +{u^{\text{UU,P2}}_{l,n}}
-\beta_1
x^{\text{Pin}}_{n,1}
-
\beta_0
d^{\text{UU}}_{l,n},\label{eq:admm_step1_slack}\\
\text{C14b}\hspace{-1mm}:&
\phi^{\text{UU}}_{l,n} +{u^{\text{UU,P2}}_{l,n}}
\hspace{-1mm}-\hspace{-1mm}
\beta_0
d^{\text{UU}}_{l,n}
-
\beta_1
x^{\text{Pin}}_{n,1}
\le
\epsilon^{\text{UU}}_{l,n},\notag\\
\text{C15a}\hspace{-1mm}:&
\hspace{-1mm}-\hspace{-1mm}
\epsilon^{\text{SI}}_{i,n,m}
\hspace{-1mm}\le\hspace{-1mm}
\phi^{\text{SI}}_{i,n,m} + \widetilde{u^{\text{SI,P2}}_{i,n,m}}
\hspace{-1mm}-\hspace{-1mm}
\beta_0
d^{\text{SI}}_{i,n,m}
-\hspace{-1mm}
\beta_1
x^{\text{Pin}}_{n,m}\hspace{-1mm}-\hspace{-1mm}\beta_1 x^{\text{Pin}}_{i,1},\notag\\
\text{C15b}:\hspace{-1mm}&
\phi^{\text{SI}}_{i,n,m}\hspace{-1mm} + \hspace{-1mm}\widetilde{u^{\text{SI,P2}}_{i,n,m}}
\hspace{-1mm}-\hspace{-1mm}
\beta_0
d^{\text{SI}}_{i,n,m}
\hspace{-1mm}-\hspace{-1mm}
\beta_1
x^{\text{Pin}}_{n,m}\hspace{-1mm}-\hspace{-1mm}\beta_1 x^{\text{Pin}}_{i,1}
\hspace{-1mm}\le\hspace{-1mm}
\epsilon^{\text{SI}}_{i,n,m},\notag
\end{align}
where
$
\widetilde{u^{\text{DU,P2}}_{k,n,m}}
=
\sum_{q=1}^{m-1}\Delta \phi_{n,q}
+
u^{\text{DU,P2}}_{k,n,m}$ and $
\widetilde{u^{\text{SI,P2}}_{i,n,m}}
=
\sum_{q=1}^{m-1}\Delta \phi_{n,q}
+
u^{\text{SI,P2}}_{i,n,m}.
$
 Then, the MM method is applied to replace 
 constraints C13b, C14b, and C15b with their first-order Taylor lower bounds at the local point 
$
x^{\text{Pin}(t_{\text{A3}})}_{n,m}
$:
        \begin{align}
&d^{\text{DU}}_{k,n,m} (x^{\text{Pin}}_{n,m})
\ge f^{\text{MM}}_{\text{D,DU}}\label{eq:dist_upperB}\\
 & \triangleq d^{\text{DU}}_{k,n,m} (x^{\text{Pin}(t_{\text{A3}})}_{n,m}) 
+
\tfrac{
x^{\text{Pin}(t_{\text{A3}})}_{n,m}
-
x^{\text{DU}}_{k}
}{
d^{\text{DU}}_{k,n,m} (x^{\text{Pin}(t_{\text{A3}})}_{n,m})
}
(
x^{\text{Pin}}_{n,m}
-
x^{\text{Pin}(t_{\text{A3}})}_{n,m}
).\notag
\end{align}
Thus, any solution satisfying
$\overline{\mathrm{C13b}}$, $\overline{\mathrm{C14b}}$, and $\overline{\mathrm{C15b}}$ in \eqref{eq:BARC13-C15B}
also satisfies constraints
$\mathrm{C13b}, \mathrm{C14b}$ and $\mathrm{C15b}$, respectively.
\begin{align}
&\overline{\text{C13b}}:
\phi^{\text{DU}}_{k,n,m} + \widetilde{u^{\text{DU,P2}}_{k,n,m}}
-
\beta_0
f^{\text{MM}}_{\text{D,DU}} 
-
\beta_1
x^{\text{Pin}}_{n,m}
\le
\epsilon^{\text{DU}}_{k,n,m},\notag\\
&\overline{\text{C14b}}:
\phi^{\text{UU}}_{l,n} +{u^{\text{UU,P2}}_{l,n}}
-
\beta_0
f^{\text{MM}}_{\text{D,UU}}
-
\beta_1
x^{\text{Pin}}_{n,1}
\le
\epsilon^{\text{UU}}_{l,n},\label{eq:BARC13-C15B}\\
&\overline{\text{C15b}}:
\phi^{\text{SI}}_{i,n,m} + \widetilde{u^{\text{SI,P2}}_{i,n,m}}
-
\beta_0
f^{\text{MM}}_{\text{D,SI}}
\hspace{-1mm}-\hspace{-1mm}
\beta_1
x^{\text{Pin}}_{n,m}-\beta_1 x^{\text{Pin}}_{i,1}
\le
\epsilon^{\text{SI}}_{i,n,m},\notag
\end{align}
where $f^{\mathrm{MM}}_{\mathrm{D,UU}}$ and
$f^{\mathrm{MM}}_{\mathrm{D,SI}}$ are lower bounds of
$d^{\mathrm{UU}}_{l,n}$ and $d^{\mathrm{SI}}_{i,n,m}$, respectively, obtained similarly to
\eqref{eq:dist_upperB}; their derivations are omitted for brevity.
Then, we handle the non-convexity in $f_{\text{P1}}$ in \eqref{eq:ADMM_block3_update}. Take the first absolute square term as an example. To verify its convexity, we expand the absolute form as 
\begin{align}\label{eq:blk3-p1}
f^{\text{DU,P1}} _{k,n,m}
=&
2|
\widetilde z^{\text{DU}}_{k,n,m}
|
|
u^{\text{DU,P1}}_{k,n,m}
-
e^{-j\Phi^{\text{DU}}_{k,n,m}}
|
\cos(\Delta\theta^{\text{DU}}_{k,n,m})
d^{\text{DU}}_{k,n,m}
\notag\\& \hspace{-2mm}+|
\widetilde z^{\text{DU}}_{k,n,m}
|^2
(d^{\text{DU}}_{k,n,m})^2+|
u^{\text{DU,P1}}_{k,n,m}
-
e^{-j\Phi^{\text{DU}}_{k,n,m}}
|^2,
\end{align}
where
$\Delta\theta^{\text{DU}}_{k,n,m} = \angle \widetilde z^{\text{DU}}_{k,n,m}
-
\angle
 (u^{\text{DU,P1}}_{k,n,m}
-
e^{-j\Phi^{\text{DU}}_{k,n,m}})
.$
Note that \eqref{eq:blk3-p1} is convex, when $\cos(\Delta\theta^{\text{DU}}_{k,n,m})\ge 0$. Otherwise, replacing
its distance term with the affine lower bound in \eqref{eq:dist_upperB}. 
Thus, 
\begin{equation}
\overline{f^{\text{DU},\text{P1}}_{k,n,m}}
=
\begin{cases}
{f^{\text{DU},\text{P1}}_{k,n,m}}, & \cos(\Delta\theta^{\text{DU}}_{k,n,m})\ge 0,\\
\widetilde{f}^{\text{DU},\text{P1}}_{k,n,m}, & \cos(\Delta\theta^{\text{DU}}_{k,n,m})< 0.
\end{cases}
\end{equation}
The same procedure can be applied to the UU and SI terms. As a result, a convex surrogate of $f_{\text P1}$ is 
\begin{align}
\overline{f_{\text P1}}
\triangleq
&
\tfrac{\rho_1}{2}
\sum_{k=1}^{K}
\sum_{n=1}^{N}
\sum_{m=1}^{M}
\overline{f^{\text{DU},\text{P1}}_{k,n,m}}
+
\tfrac{\rho_1}{2}
\sum_{l=1}^{L}
\sum_{n=1}^{N}
\overline{f^{\text{UU},\text{P1}}_{l,n}}
\notag\\[-1mm]&+
\tfrac{\rho_1}{2}
\sum_{i\neq n}^{N}
\sum_{n=1}^{N}
\sum_{m=1}^{M}
\overline{f^{\text{SI},\text{P1}}_{i,n,m}} .
\end{align}
By replacing $f_{\text P1}$ with a convex
surrogate $\overline{f_{\text P1}}$ and constraints C13b, C14b, and C15b with
their conservative convex approximations, the convex
surrogate problem for \eqref{eq:ADMM_block3_update} is obtained as
\begin{align}
\min_{\substack{{x}^{\text{Pin}}_{n,m},\epsilon^{\text{DU}}_{k,n,m},\\\epsilon^{\text{UU}}_{l,n},\,\epsilon^{\text{SI}}_{i,n,m}}}
\quad
&
\overline{f_{\text P1}}
+
\tfrac{\rho_2}{2}
\sum_{k=1}^{K}
\sum_{n=1}^{N}
\sum_{m=1}^{M}
(\epsilon^{\text{DU}}_{k,n,m})^2
\\[-2mm]&
+\tfrac{\rho_2}{2}
\sum_{l=1}^{L}
\sum_{n=1}^{N}
(\epsilon^{\text{UU}}_{l,n})^2
+
\tfrac{\rho_2}{2}
\sum_{i\neq n}^{N}
\sum_{n=1}^{N}
\sum_{m=1}^{M}
(\epsilon^{\text{SI}}_{i,n,m})^2\notag
\\
\text{s.t.}
\quad
&
\text{C2-C5},\text{C13a},\overline{\text{C13b}},\text{C14a},\overline{\text{C14b}},\text{C15a},\overline{\text{C15b}}.\notag
\end{align}

\label{sol:PBs-ADMM}
\begingroup
\setlength{\textfloatsep}{6pt plus 1pt minus 1pt}
\begin{algorithm}[t]
\caption{Proposed Hierarchical BCD-Based Algorithm}
\label{alg:proposed}
\footnotesize
\begin{algorithmic}[1]
\STATE Initialize a feasible solution point to \eqref{proposed_formulation_origion}, $t_{\rm O}=0$, $T_{\max}$, and
$\epsilon_{\rm out}$.

\REPEAT
\STATE \textbf{Block 1:} Alternately update
$\{\bm w_{k|\mathrm{ext}}\}$ via
\eqref{eq:mmse_receiver}--\eqref{eq:wmmse_tx_qp} and
$\{\bm v_{l|\mathrm{ext}},p_l^{\mathrm U}\}$; with
$\{\bm w_{k|\mathrm{ext}}\}$ fixed, update
$\{\bm v_{l|\mathrm{ext}}\}$ via \eqref{eq:mmse_combiner} and
$\{p_l^{\mathrm U}\}$ via
\eqref{eq:fixed_point_update}.
Repeat until Block 1 convergence.
\STATE \textbf{Block 2:} Repeat the following updates until convergence:
update $\{\alpha_k\}$ and $\{\nu_k\}$ using
\eqref{eq:FP_alpha}--\eqref{eq:FP_nu}; update $\{s_n\}$ and
$\{\eta_{n,m}\}$ by solving \eqref{eq:FP-SP1}; and update the PA locations
$\{x_{n,m}^{\mathrm{Pin}}\}$ using the ADMM procedure in
\eqref{eq:ADMM_block1_update}--\eqref{eq:dual_p2}.

\STATE $t_{\rm O}\leftarrow t_{\rm O}+1$.

\UNTIL{$t_{\rm O}\geq T_{\max}$ or the relative objective value change is no
greater than $\epsilon_{\rm out}$}
\STATE \textbf{Return} the optimized solution.
\end{algorithmic}
\end{algorithm}\endgroup\vspace{-2mm}
 The proposed hierarchical BCD method is summarized in
Algorithm~\ref{alg:proposed}. Each block is updated leveraging closed-form
expressions, tractable convex subproblems solved by standard convex
optimization solvers, or low-complexity iterative procedures.
The overall computational complexity \cite{boyd2004convex} is
\begin{align}
\mathcal{O}\Big(T_{\rm O}\Big[
&T_{\rm B1}\big(T_{\rm W}N^3+T_{\rm U}LN^3\big)+T_{\rm B2}\big(
K^2NM+T_{\rm MM}D_1^{3.5}\notag\\
&
+T_{\rm ADMM}(D_Z^{3.5}+D_X^{3.5}+QT_\phi)
\big)\Big]\Big),
\end{align}
where
$D_1=2NM+N+N(N-1)M$, $Q=KNM+LN+N(N-1)M$,
$D_Z=2Q$, and $D_X=NM+Q$.
Here, $T_{\rm O}$, $T_{\rm B1}$, and $T_{\rm B2}$ denote the iteration
counts of the outer BCD loop and Blocks~1 and~2, respectively; $T_{\rm W}$
and $T_{\rm U}$ are the WMMSE and uplink combiner--power iteration counts;
$T_{\rm MM}$ and $T_{\rm ADMM}$ denote the MM and ADMM iteration counts,
and $T_\phi$ is the maximum number of Newton iterations per phase update.
Thus, the proposed algorithm exhibits polynomial computational
complexity in the system dimensions for fixed iteration counts and
yields a high-quality suboptimal solution to the original problem.\label{sol:con_proof}

\section{Simulation Results}

\subsection{Simulation Setup}
This section evaluates the performance of the proposed FD multi-waveguide
PA system. To characterize the PA-induced phase loading in~\eqref{eq:add_phase}, motivated
by the millimeter-scale polytetrafluoroethylene (PTFE) dielectric-waveguide
and PA geometries reported in~\cite{Li2026Geometry,VU2021PTFE} and the
localized dielectric-coupler model in~\cite{Wang2025PASS}, we set
$\epsilon_{r,\mathrm{PA}}=2.1$, $\epsilon_{r,\mathrm{air}}=1$. We assume identical PA geometries \cite{Wang2025PASS}, with $D_{\mathrm{PA}}=12$~mm,
$(w_y,w_z)=(3,2)$~mm, and $A_{\mathrm{ov}}=1~\mathrm{mm}^2$.
Accordingly, $A_{\mathrm{eff}}=\pi w_yw_z=18.85~\mathrm{mm}^2$ and
$A_{\mathrm{ov}}/A_{\mathrm{eff}}\approx 0.053$.
Unless otherwise specified, we consider $N=4$ waveguides, each equipped with
$M=4$ PAs, serving $K=2$ downlink users and $L=2$ uplink users. The waveguides are positioned
at a height of $d=3$~m above the user plane with a separation of $2$~m between adjacent waveguides. Each PA is positioned within
$x_{n,m}^{\mathrm{Pin}}\in[0,20]$~m, subject to a minimum separation of
$\delta x=\lambda_0/2$. The UL and DL users are randomly distributed within a \(20\,\text{m}\times20\,\text{m}\) region beneath the waveguides. The carrier frequency is $f_c=28$~GHz, while the
effective refractive index and dielectric loss tangent are
$n_{\mathrm{eff}}=1.4$ and $\tan(\delta_e)=2\times10^{-4}$, respectively.
The total nominal DL transmit-power budget is set to $35$~dBm and equally
allocated among all waveguides, i.e.,
$P^{\mathrm{D}}_{\max,n}=P^{\mathrm{D}}_{\max}/N$, $\forall n$.
The UU$_l$ transmit-power budget is $P_{\max,l}^{\text{U}}=10$~dBm.
The noise powers at the BS and DL users are
$\sigma_{\mathrm{BS}}^2=-90$~dBm and
$\sigma_{\mathrm{DU}}^2=-100$~dBm, respectively. We further set
$\mu_k=1,\forall k$, $\rho_{\mathrm{ANA}}=10^{-3}$,
$R_{\min}^{\text{UU}}=1$~bit/s/Hz, and
$\eta_{\max}^{\text{UL}}=1$.

%\begin{table}[t]
%\centering
%\caption{Default simulation parameters.}
%\label{tab:simulation_parameters}
%\renewcommand{\arraystretch}{1.1}
%\setlength{\tabcolsep}{3pt}
%\begin{tabular}{p{0.62\columnwidth}p{0.27\columnwidth}}
%\hline
%\textbf{Parameter} & \textbf{Value} \\
%\hline
%
%Number of dielectric waveguides, $N$
%& $4$ \\
%
%Number of PAs per waveguide, $M$
%& $4$ \\
%
%Number of DL users, $K$
%& $2$ \\
%
%Number of UL users, $L$
%& $2$ \\
%
%Inter-waveguide distance,
%$y_{n+1}^{\mathrm{FP}}-y_n^{\mathrm{FP}}$
%& $2$ m \\
%
%Waveguide height, $d$
%& $3$ m \\
%
%PA moving range, $x_{n,m}^{\mathrm{Pin}}$
%& $[0,20]$ m \\
%
%Minimum PA spacing, $\Delta_x$
%& $\lambda_0/2$ \\
%
%Carrier frequency, $f_c$
%& $28$ GHz \\
%
%Effective refractive index, $n_{\mathrm{eff}}$
%& $1.4$ \\
%
%Effective dielectric loss tangent, $\tan(\delta_e)$
%& $2\times10^{-4}$ \\
%
%Maximum DL transmit power, $P_{\mathrm{D,max}}$
%& $30$ dBm \\
%
%Maximum UL transmit power, $P_{\mathrm{U,max}}$
%& $20$ dBm \\
%
%Noise power at the BS, $\sigma_{\mathrm{BS}}^2$
%& $-90$ dBm \\
%
%Noise power at the DL users, $\sigma_{\mathrm{DU}}^2$
%& $-100$ dBm \\
%
%Residual SI factor, $\rho_{\mathrm{ANA}}$
%& $10^{-3}$ \\
%
%Minimum UL rate requirement, $R_{\mathrm{UU,min}}$
%& $1$ bit/s/Hz \\
%
%Maximum UL collection ratio, $\eta_{\mathrm{UL,max}}$
%& $1$ \\
%
%PA relative permittivity, $\epsilon_{r,\mathrm{PA}}$
%& $2.1$ \\
%
%Modal waists, $(w_1,w_2)$
%& $(3,2)$ mm  \\
%
%Effective overlap area, $A_{\mathrm{ov}}$
%& $1$ mm$^2$ \\
%
%PA coupling length, $D_{\mathrm{PA}}$
%& $12$ mm \\
%
%\hline
%\end{tabular}
%\end{table}
\vspace{-2mm}

\subsection{Accuracy of the PA-Induced Phase-Shift Approximation}
\begin{figure}[t] 
  \centering
  \includegraphics[width=\columnwidth]{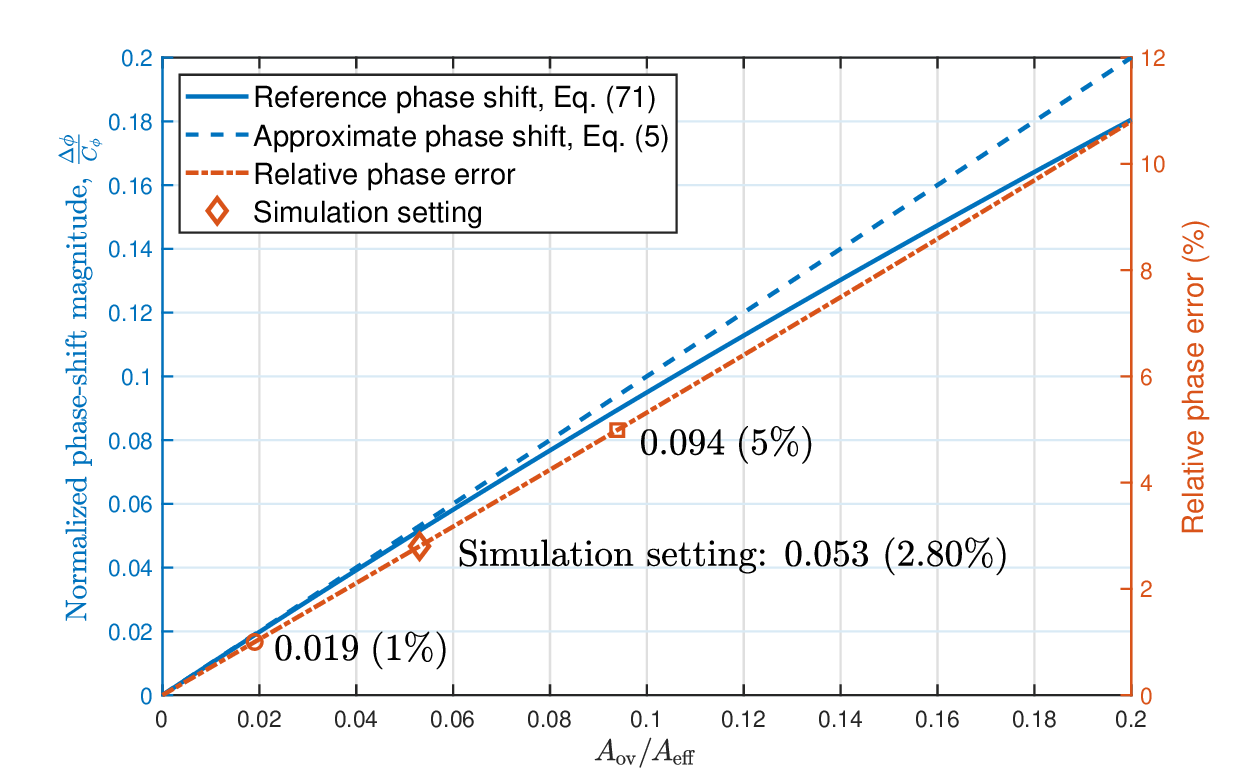}\vspace{-0mm}
\caption{Relative error of the closed-form approximation in
\eqref{eq:add_phase} compared with the phase shifts obtained from
\eqref{eq:sigma_integral}, versus $A_{\mathrm{ov}}/A_{\mathrm{eff}}$.}
  \label{fig:eq5_validation}\vspace{-3mm}
\end{figure}
Fig.~\ref{fig:eq5_validation} compares the approximate phase shift obtained from \eqref{eq:add_phase} with the reference phase shift  $\Delta\phi_{\mathrm{ref}}$   computed using \eqref{eq:sigma_integral}, with both normalized by
$C_{\phi}=-\tfrac{\pi(\epsilon_{r,\mathrm{PA}}- \epsilon_{r,\mathrm{air}})D_{\mathrm{PA}}}{\lambda_0 n_{\mathrm{eff}}}
$. The approximation accuracy is quantified by the relative phase error
$
\epsilon_{\phi}
=
\tfrac{
|\Delta\phi-\Delta\phi_{\mathrm{ref}}|
}{
|\Delta\phi_{\mathrm{ref}}|
}
\times 100\%.$ 
 The phase shifts converge as
$A_{\mathrm{ov}}/A_{\mathrm{eff}}$ decreases, supporting the small-footprint approximation. At the adopted $A_{\mathrm{ov}}/A_{\mathrm{eff}}=0.053$, the 
error is $2.8\%$, indicating sufficient accuracy.

\vspace{-2mm}
\subsection{Performance Evaluation and Design Insights}

We consider four benchmark schemes: i) \textbf{Conventional FD BS}, which is located at the center of the waveguide feed points and equipped with $N$ antennas. The downlink beamformers, uplink combiners, and transmit powers are optimized.
ii) \textbf{Fixed-PA}, which follows the proposed optimization framework but with randomly generated
feasible PA locations fixed during optimization; iii) \textbf{Fixed-WG}, which follows
the proposed optimization framework with randomly generated feasible waveguide-mode
assignments fixed during optimization; and iv) \textbf{Phase-Ignorant}, which follows
the proposed optimization framework while neglecting PA-induced phase loading. Its resulting
allocations are re-evaluated leveraging the actual phase-aware channel model. 
In particular, all schemes are evaluated over 100 independently generated user-location realizations. The proposed scheme is feasible in all realizations, whereas the baselines are infeasible in some realizations, depending on the setting. For each scheme, the average sum rate (SR) is computed over its feasible realizations. Missing markers indicate that no feasible realization was obtained. For \textbf{Fixed-PA} and \textbf{Fixed-WG}, the shaded regions indicate the minimum and maximum average SRs obtained across multiple randomly generated feasible configurations.

\subsection{Downlink SR versus Transmit Power Budget}
\label{subsec:power_results}

\begin{figure}[t]
    \centering
    \includegraphics[width=\columnwidth]{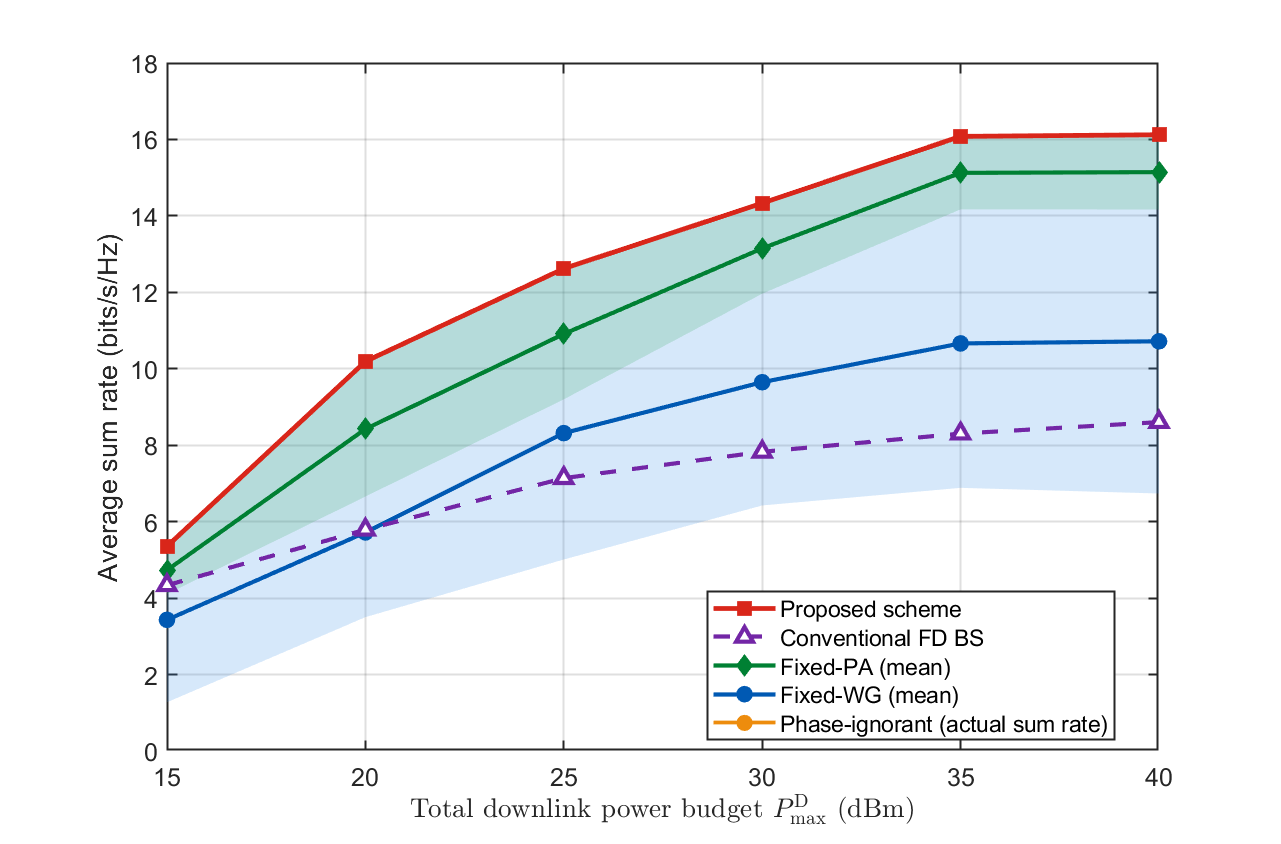}\vspace{-2mm}
    \caption{Average downlink SR versus the maximum downlink
    transmit power budget $P_{\mathrm{D}}^{\max}$.}\vspace{-2mm}
    \label{fig:wsr_vs_dl_power}\vspace{-3mm}
\end{figure}

Fig.~\ref{fig:wsr_vs_dl_power} shows the average downlink SR versus
the maximum downlink transmit-power budget $P_{\mathrm{D}}^{\max}$. The \textbf{proposed scheme} consistently achieves the highest
WSR. Compared with a \textbf{conventional FD BS}, PASS can relocate its
radiation points along the waveguides closer to the users, thereby
reducing the access-link path loss and improving the received signal
strength.
Compared with \textbf{Fixed-PA}, the \textbf{proposed scheme} enhances coherent signal combining by optimizing PA locations according to the user distribution. More importantly, the larger gap between the \textbf{proposed scheme} and the \textbf{Fixed-WG} scheme indicates that adaptive waveguide mode selection provides greater gain than PA location optimization under the considered settings. Indeed, waveguide mode selection determines the spatial distribution of transmission and reception points, thereby jointly influencing the desired-link gains and residual SI coupling. In contrast, PA relocation only refines the antenna geometry within a fixed waveguide mode assignment.
The wide shaded region of \textbf{Fixed-WG} further indicates that its performance is sensitive to the selected waveguide modes, highlighting the benefit of adaptive mode selection. Moreover, SR gains diminish at high transmit-power budgets across all schemes. For PASS, the finite PA positioning range and hardware-dependent responses limit further improvements in desired-signal combining and SI suppression that could be provided by an additional transmit power budget, while uplink QoS constraints restrict the achievable trade-off.
Finally, the \textbf{Phase-Ignorant} scheme is generally infeasible when evaluated under the phase-aware channel model, demonstrating that neglecting cumulative PA-induced phase
shifts can lead to uplink QoS violations.

\subsection{Adaptive Waveguide-Mode and PA Deployment}\label{subsec:deployment_results}

\begin{figure*}[t]%\vspace{-3mm}
    \centering
    \begin{minipage}[t]{\columnwidth}
        \centering
        \includegraphics[width=\linewidth]{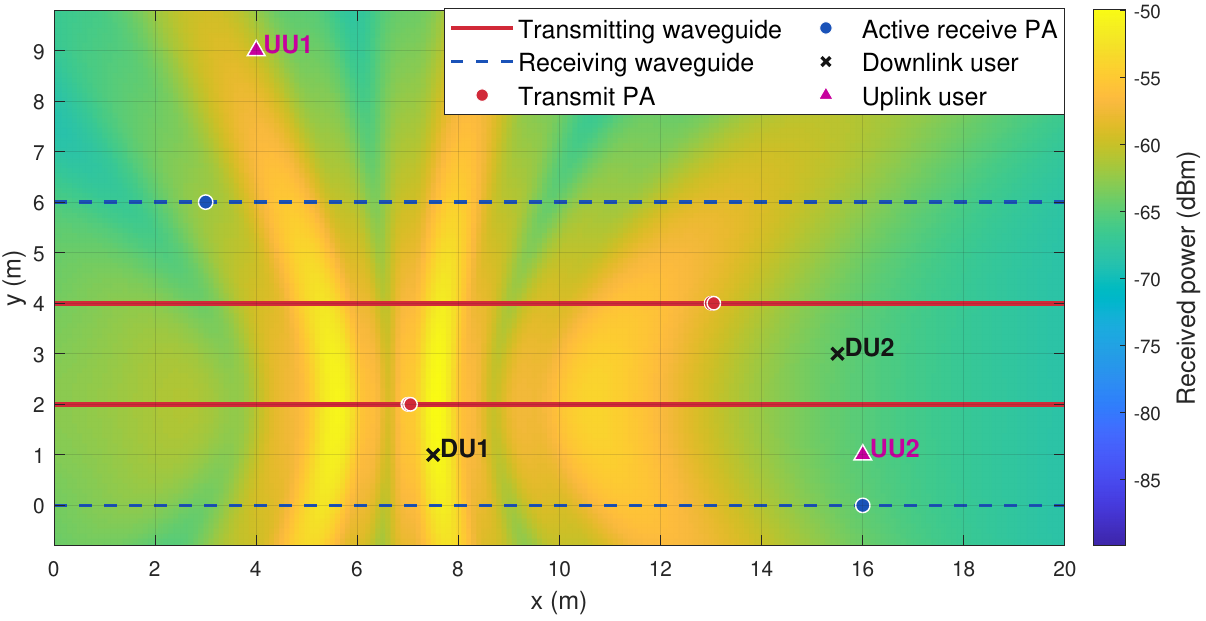}
       \vspace{-1.5mm}
        \footnotesize (a) Power distribution generated
by the downlink stream intended for DU$_1$.
    \end{minipage}
    \hfill
    \begin{minipage}[t]{\columnwidth}
        \centering
        \includegraphics[width=\linewidth]{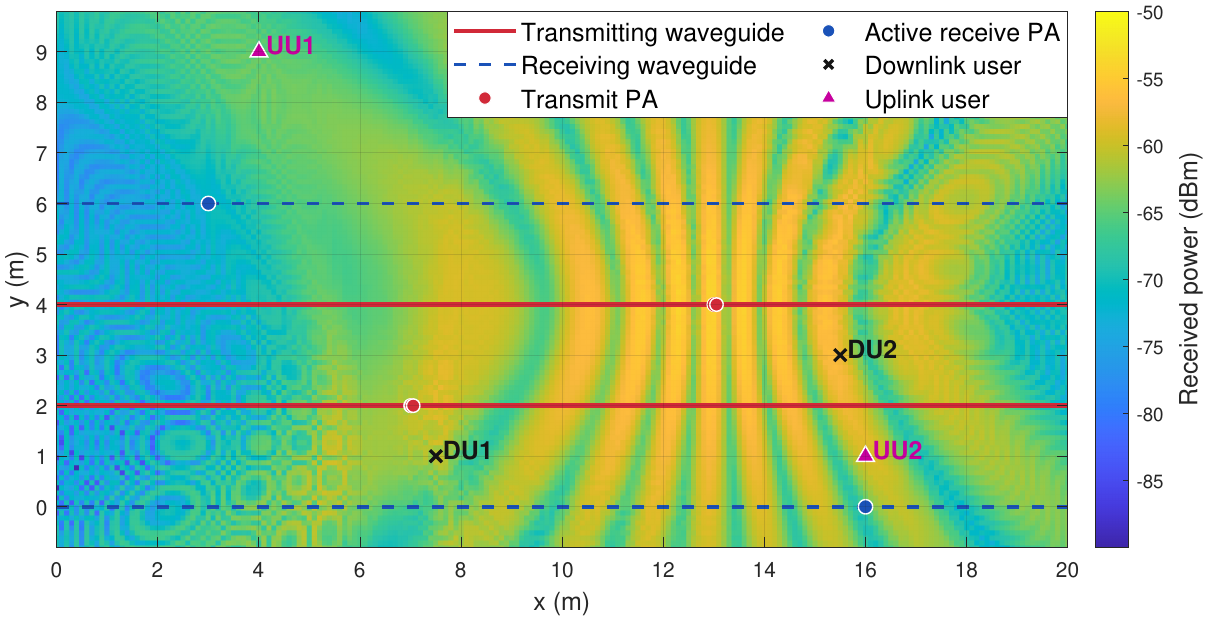}
       %\vspace{-1.5mm}
        \footnotesize (b) Power distribution generated by
the downlink stream intended for DU$_2$.
    \end{minipage}
    \caption{Optimized PA deployment and spatial received-power distributions for
a representative user-location realization.}
    \label{fig:optimized_deployments}\vspace{-3mm}
\end{figure*}

Fig.~\ref{fig:optimized_deployments} illustrates the optimized PASS configuration and the corresponding spatial received-power distributions for a representative user-location realization. Each DU receives relatively high power from its intended downlink stream, while the leakage from the stream intended for the other DU is effectively suppressed. This spatial selectivity is achieved by jointly optimizing the PA locations and digital beamformers to shape the effective downlink channels, rather than simply positioning the PAs closer to the user.

\subsection{Impact of Residual Self-Interference}\label{subsec:residual_si_results}

\begin{figure}[t]
    \centering\vspace{-2mm}
    \includegraphics[width=\columnwidth]{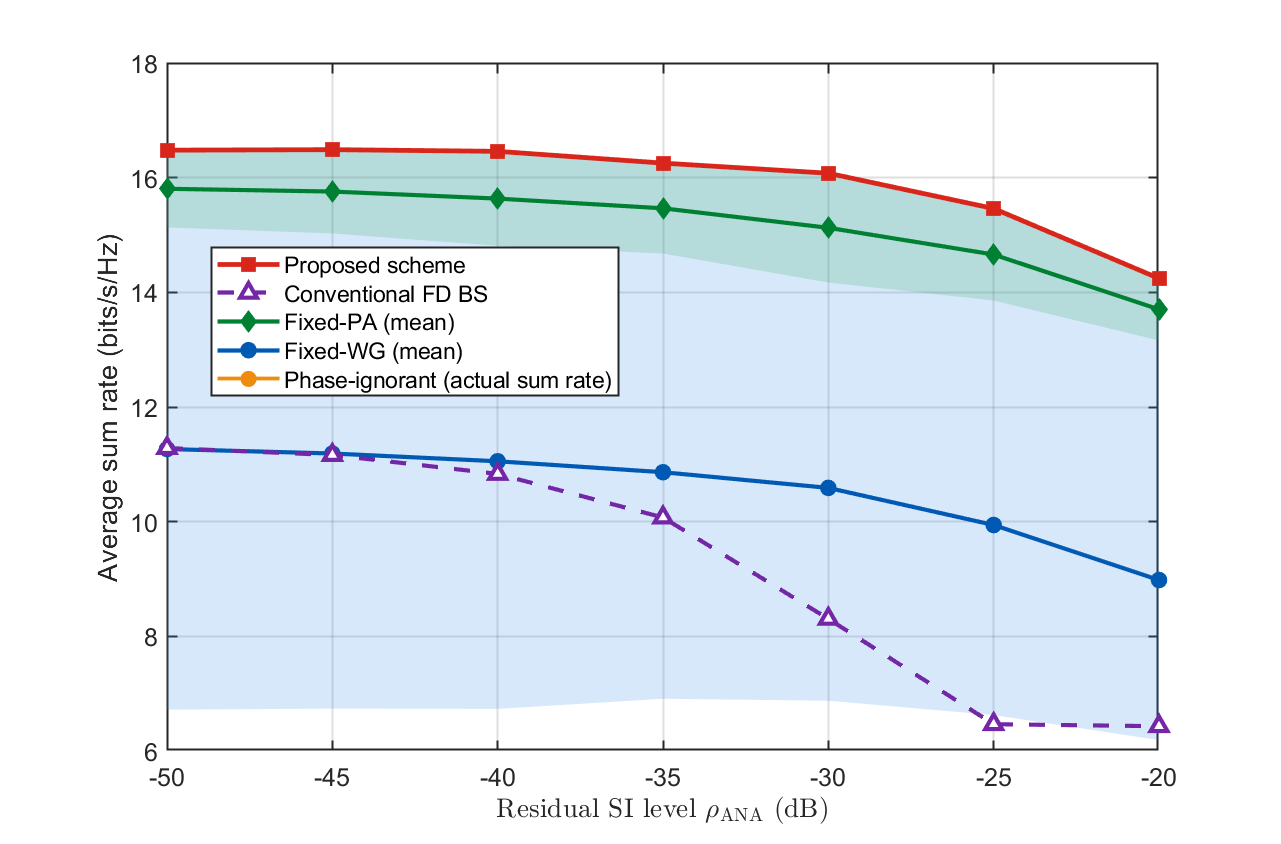}\vspace{-2mm}
    \caption{Average downlink SR versus the residual SI level, i.e.,
    $10\log_{10}(\rho_{\mathrm{ANA}})$ (dB).}
    \label{fig:wsr_vs_residual_si}\vspace{-3mm}
\end{figure}

Fig.~\ref{fig:wsr_vs_residual_si} shows the average downlink SR versus the residual SI level $\rho_{\mathrm{ANA}}$. As $\rho_{\mathrm{ANA}}$ increases, stronger residual SI imposes tighter restrictions on downlink transmission to maintain the required uplink QoS, thereby reducing the downlink SR. The \textbf{proposed scheme} consistently achieves the highest SR and exhibits less performance degradation than the \textbf{conventional FD BS}. This greater robustness is attributed to the joint adaptation of waveguide modes and PA positions, which provides additional spatial DoFs to strengthen the desired links while reducing SI coupling. In contrast, the \textbf{conventional FD BS} relies primarily on beamforming to manage SI, without the flexibility to reconfigure its transmit-receive antenna geometry.

\begin{figure}[t]\vspace{-2mm}
    \centering
    \includegraphics[width=\columnwidth]{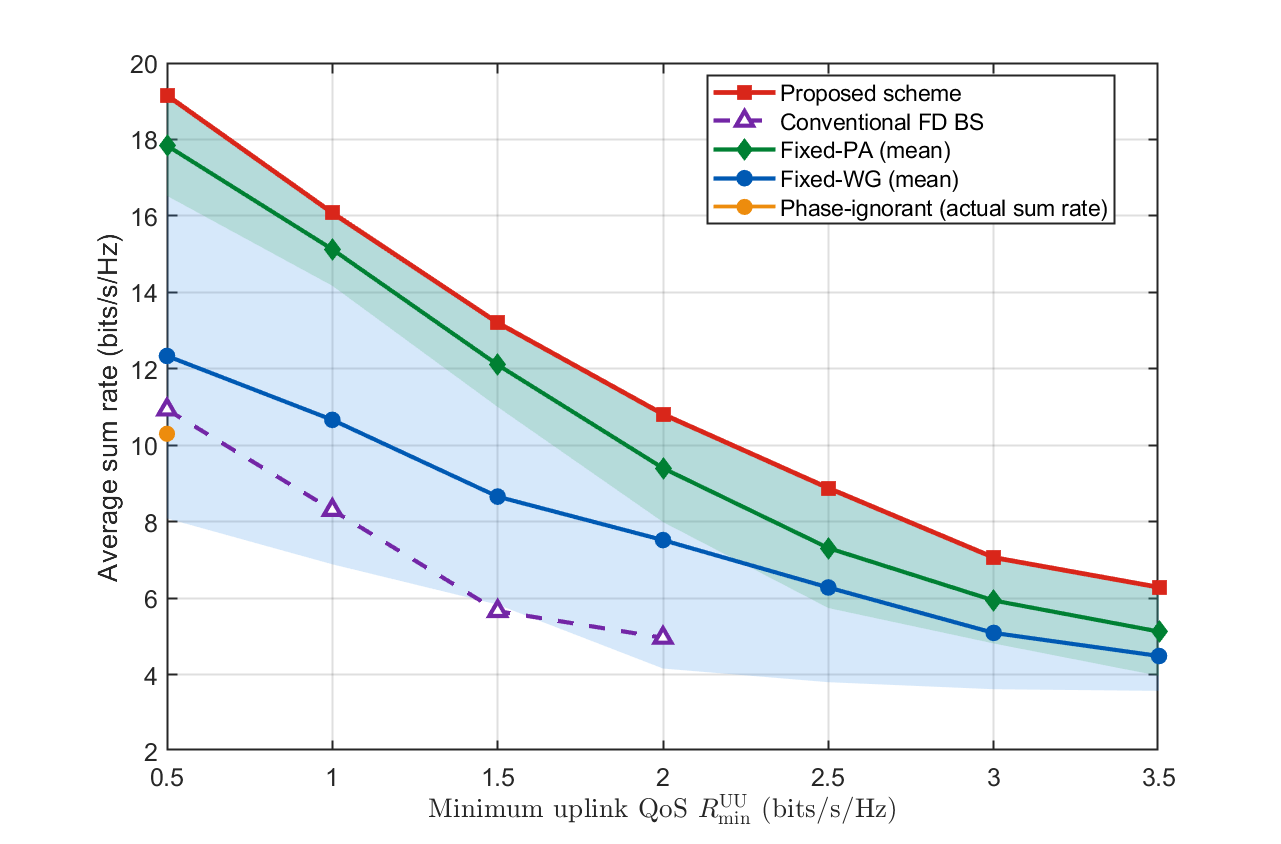}\vspace{-2mm}
    \caption{Average downlink SR versus the minimum uplink rate, $R^{\mathrm{UU}}_{\min}$.}
    \label{fig:wsr_vs_ul_qos}\vspace{-3mm}
\end{figure}

Fig.~\ref{fig:wsr_vs_ul_qos} compares the average downlink SR of different schemes as the minimum uplink rate requirement $R_{\min}^{\mathrm{UU}}$ increases. The \textbf{PASS-based schemes} remain feasible under more stringent uplink QoS requirements than the \textbf{conventional FD BS}. This improved feasibility stems from the additional spatial DoFs provided by adaptive PA configurations, which enable PASS to enhance the desired uplink channels while mitigating residual SI. Consequently, PASS can accommodate higher uplink rate requirements while maintaining downlink communication performance.

\label{sol:PBs-ETAS}

\section{Conclusion}
This paper developed a multiuser FD-PASS design that accounts for in-waveguide
attenuation and cumulative PA-induced phase shifts. We proposed   a low-complexity hierarchical
WMMSE--FP--MM--ADMM algorithm to jointly optimize the waveguide modes, PA
locations and radiation/collection coefficients, transceivers, and uplink
powers to maximize the downlink WSR subject to uplink QoS and power constraints.
Numerical results highlighted two key insights. First, adaptive waveguide-mode assignment and PA locations allow PASS to dynamically exploit its spatial DoFs to better separate downlink and uplink operations and mitigate residual SI.  Second, cumulative PA-induced phase loading is essential to accurate channel characterization and neglecting it can lead to designs that fail to satisfy uplink QoS.
\appendix\label{appA}
%\subsection{Phase Shift Induced by the PA} 
%When a PA operates as a non-contact coupler near a dielectric waveguide, it extracts power from and perturbs the guided mode. 
Under the weak-coupling and symmetric-coupler assumptions, the PA-induced perturbation is modeled using coupled-mode theory (CMT) \cite{johnson2002perturbation,okamoto2021fundamentals}. Let \(A(x)\) and \(B(x)\) denote the modal amplitudes of the main waveguide and PA,  respectively, with propagation constants \(\beta_1\) and \(\beta_2\), respectively. The coupled-mode equations are
\begin{align}
\tfrac{dA(x)}{dx}
&=
-j\kappa B(x) e^{-j\Delta\beta x}
+j\sigma A(x),\notag\\
\tfrac{dB(x)}{dx}
&=
-j\kappa A(x) e^{j\Delta\beta x}
+j\sigma B(x),\notag\\
\sigma &= - \tfrac{\omega\epsilon_0}{4P_1}
{\scriptsize\iint}
\bigl(
\Delta\epsilon_r 
\bigr)
|\bm E_1(y,z)|^2
\,dy\,dz,
\label{eq:Xp}
\end{align}
where \(\kappa\geq 0\) is the mutual coupling coefficient, 
\(\Delta\beta=\beta_1-\beta_2\) is the propagation-constant mismatch, 
and \(\sigma\) is the self-coupling coefficient induced by the PA. 
The real part of \(\sigma\) represents phase loading per unit length, 
whereas its imaginary part represents loss. Moreover, 
\(P_1\) is the guided power in the main waveguide, 
\(\bm E_1(y,z)\) is its transverse electric-field distribution, and
$
\Delta\epsilon_r(y,z)
=
\epsilon_{r,\mathrm{PA}}(y,z)
-
\epsilon_{r,\mathrm{air}}(y,z)
$
denotes the relative-permittivity perturbation caused by introducing the PA, where \(\epsilon_{r,\mathrm{PA}}\) and \(\epsilon_{r,\mathrm{air}}\) are the relative-permittivity distributions with and without the PA, respectively. Here, \(\omega=2\pi f_c\) is the angular frequency and \(\epsilon_0\) is the free-space permittivity.

For the phase-loading model used in this work, we consider the
phase-matched case \(\Delta\beta=0\). Under the adopted
weak-coupling through-mode approximation, the PA-induced phase
accumulation is determined by the real part of the self-coupling
coefficient. Over the PA interaction length \(D_{\mathrm{PA}}\),
the resulting through-mode phase shift is 
\begin{equation}\small
\Delta\phi_{\text{through}}
=\Re\{\sigma \}D_{\text{PA}}.
\label{eq:sigma_integral}
\end{equation}
Under the CMT convention in \eqref{eq:Xp}, the signed phase shift
$\Delta\phi_{\mathrm{through}}$ corresponds to the signal multiplier
$e^{+j\Delta\phi_{\mathrm{through}}}$. 

 Let $(y_c,z_c)$ denote the
center of the perturbation region. When the PA footprint is
small compared with the transverse variation scale of the
guided mode, the modal intensity is approximately uniform
within the perturbation region. Therefore,
\begin{equation}\label{eq:E_ing}\small
{\scriptsize\iint}_{A_{\mathrm{ov}}}
\Delta\epsilon_r
|\bm{E}_1(y,z)|^2
\,\mathrm{d}A
\approx
\Delta\epsilon_r
|\bm{E}_1(y_c,z_c)|^2
A_{\mathrm{ov}},
\end{equation}
where  $A_{\text{ov}}$ is the footprint area overlapping the field. Assuming transverse PA alignment with the fundamental mode’s high-intensity region, its intensity is modeled by the elliptical Gaussian profile, i.e., $
|\bm{E}_1(y,z)|^2
\approx
|\bm{E}_{1,\max}|^2
\exp(
-\tfrac{(y-y_0)^2}{w_y^2}
-\tfrac{(z-z_0)^2}{w_z^2}
)$,
where $(y_0,z_0)$ denotes the modal center and $w_y$ and
$w_z$ are the transverse modal beam waists \cite{snyder1983optical}. Under this approximation,
the effective modal area is defined as
$
A_{\mathrm{eff}}
\triangleq
\tfrac{
\iint |\bm{E}_1(y,z)|^2\,\mathrm{d}A
}{
|\bm{E}_{1,\max}|^2
}
=
\pi w_yw_z.
$
Using the guided-mode power normalization
$
P_1
\approx
\tfrac{1}{2}
\epsilon_0 n_{\mathrm{eff}} c
\iint |\bm{E}_1(y,z)|^2\,\mathrm{d}A,
$
and the high-field alignment assumption
$|\bm{E}_1(y_c,z_c)|^2\approx |\bm{E}_{1,\max}|^2$, the local
modal intensity can be expressed as
\begin{equation}\label{eq:E_ra}\small
|\bm{E}_1(y_c,z_c)|^2
\approx
|\bm{E}_{1,\max}|^2
\approx
\tfrac{2P_1}
{\epsilon_0 n_{\mathrm{eff}}cA_{\mathrm{eff}}}.
\end{equation}
%It is worth noting that \eqref{eq:E_ing} and \eqref{eq:E_ra} rely on two distinct conditions. Equation \eqref{eq:E_ing} follows from the small-footprint condition, whereas \eqref{eq:E_ra}  follows from the transverse high-field alignment and the Gaussian effective-mode representation.
Substituting \eqref{eq:E_ing} and \eqref{eq:E_ra}  into \eqref{eq:sigma_integral}, we obtain
$%\begin{equation}\small
\Re\{\sigma\}
\approx
-\tfrac{\pi\Delta\epsilon_r}
{\lambda_0n_{\mathrm{eff}}}
\tfrac{A_{\mathrm{ov}}}
{A_{\mathrm{eff}}}
$%\end{equation}
and Eq. \eqref{eq:add_phase}.

\label{app:add_phase}

\bibliographystyle{IEEEtran}
\bibliography{pinching_ant_FD}

@STRING{com = {IEEE Trans. Commun.} }

@STRING{globecom = {Proc. IEEE Global Telecommun. Conf.} }

@STRING{icc      = {Proc. IEEE Intern. Commun. Conf.}}

@STRING{sel = {IEEE J. Select. Areas Commun.}}

@STRING{sip = {IEEE Trans. Signal Process.} }

@STRING{wcom= {IEEE Trans. Wireless Commun.} }

@techreport{ITURM2160,
  author      = {{ITU-R}},
  title       = {Framework and Overall Objectives of the Future
                 Development of {IMT} for 2030 and Beyond},
  institution = {International Telecommunication Union
                 Radiocommunication Sector},
  type        = {Recommendation},
  number      = {M.2160-0},
  address     = {Geneva, Switzerland},
  month       = nov,
  year        = {2023},
  url         = {https://www.itu.int/dms_pubrec/itu-r/rec/m/
                 R-REC-M.2160-0-202311-I!!PDF-E.pdf},
  note        = {Accessed: Sep. 21, 2026}
}

@article{hu2026spatially,
  title={Spatially Reconfigurable Pinching-Antenna Systems: Experimental Validation and {ISAC} Applications},
  author={Hu, Shaokang and Zhao, Ruotong and Wang, Qigejian and Atakaramians, Shaghik and Ng, Derrick Wing Kwan and Yuan, Jinhong},
  journal={arXiv preprint arXiv:2610.04769},
  year={2026}
}

@article{Rangan2014mmWave,
  author  = {S. Rangan and T. S. Rappaport and E. Erkip},
  title   = {Millimeter-Wave Cellular Wireless Networks:
             Potentials and Challenges},
  journal = {Proceedings of the IEEE},
  volume  = {102},
  number  = {3},
  pages   = {366--385},
  year    = {2014},
  doi     = {10.1109/JPROC.2014.2299397}
}

@article{hu2021robustSecureSumRateIRS,
  author  = {Hu, Shaokang and Wei, Zhiqiang and Cai, Yuanxin and Liu, Chang and Ng, Derrick Wing Kwan and Yuan, Jinhong},
  title   = {Robust and Secure Sum-Rate Maximization for Multiuser {MISO} Downlink Systems with Self-Sustainable {IRS}},
  journal = {IEEE Trans. Commun.},
  volume  = {69},
  number  = {10},
  pages   = {7032--7049},
  month   = {Oct.},
  year    = {2021},
  doi     = {10.1109/TCOMM.2021.3097140}
}

@article{new2025fluidantennas,
  author  = {New, W. K. and others},
  title   = {Fluid antenna systems: Redefining reconfigurable wireless communications},
  journal = {IEEE J. Sel. Areas Commun.},
  volume  = {43},
  number  = {3},
  pages   = {800--818},
  month   = {Mar.},
  year    = {2025}
}

@article{zhu2024movableantennas,
  author  = {Zhu, L. and others},
  title   = {Movable antennas for wireless communication: Opportunities and challenges},
  journal = {IEEE Commun. Mag.},
  volume  = {62},
  number  = {6},
  pages   = {114--121},
  month   = {Jun.},
  year    = {2024}
}

@article{zhang2021active,
  author={Zhang, Zijian and Dai, Linglong and Chen, Xibi and Liu, Changhao and Yang, Fan and Schober, Robert and Poor, H. Vincent},
  journal=com, 
  title={Active {RIS} vs. Passive {RIS}: Which Will Prevail in {6G}?}, 
  year={Mar. 2023},
  volume={71},
  number={3},
  pages={1707-1725},
  doi={10.1109/TCOMM.2022.3231893}}

@ARTICLE{6832464,
  author={Sabharwal, Ashutosh and Schniter, Philip and Guo, Dongning and Bliss, Daniel W. and Rangarajan, Sampath and Wichman, Risto},
  journal=sel, 
  title={In-Band Full-Duplex Wireless: Challenges and Opportunities}, 
  year={Jun. 2014},
  volume={32},
  number={9},
  pages={1637-1652},
  doi={10.1109/JSAC.2014.2330193}}

@ARTICLE{Wang2025PASS,
  author={Wang, Zhaolin and Ouyang, Chongjun and Mu, Xidong and Liu, Yuanwei and Ding, Zhiguo},
  journal={IEEE Trans. Commun.}, 
  title={Modeling and Beamforming Optimization for Pinching-Antenna Systems}, 
  year={Oct. 2025},
  volume={73},
  number={12},
  pages={13904-13919},
  doi={10.1109/TCOMM.2025.3621049}}

@ARTICLE{11263923,
  author={Xu, Xiaoxia and Mu, Xidong and Wang, Zhaolin and Liu, Yuanwei and Nallanathan, Arumugam},
  journal={IEEE Trans. Commun.}, 
  title={Pinching-Antenna Systems ({PASS}): Power Radiation Model and Optimal Beamforming Design}, 
  year={Nov. 2025},
  volume={74},
  number={},
  pages={2160-2175},
  doi={10.1109/TCOMM.2025.3636083}}

@article{VU2021PTFE,
  author  = {Thanh-Luan Vu and St{\'e}phane Barlerin and Yves Stricot
             and Ronan Sauleau and Mauro Ettorre
             and David Gonz{\'a}lez-Ovejero},
  title   = {Experimental Characterization of Bending Effects for
             Solid and Hollow Dielectric Waveguides at {V}-Band},
  journal = {Scientific Reports},
  volume  = {11},
  pages   = {20679},
  year    = {2021},
  doi     = {10.1038/s41598-021-00187-9}
}

@ARTICLE{blockage,
  author={Zhao, Ruotong and Hu, Shaokang and Mishra, Deepak and Ng, Derrick Wing Kwan},
  journal={IEEE Trans. Mob. Comput.}, 
  title={Robust and Secure Blockage-Aware Pinching Antenna-Assisted Wireless Communication}, 
  volume={25},
  number={10},
  pages={17547--17565},
  year = {Oct. 2026}
  }

@inproceedings{Zhu2026FDWPCN,
  author    = {Zhu, Zhengyu and Li, Yixuan and Hua, Meng and
               Huang, Chongwen and Zheng, Beixiong and Xu, Mingliang},
  title     = {Reconfigurable Pinching-Antenna Assisted Full-Duplex
               Wireless-Powered Networks Under Self-Interference},
  booktitle = {Proc. IEEE Int. Conf. Commun. Workshops (ICC Workshops)},
  address   = {Glasgow, U.K.},
  month     = {May},
  year      = {2026},
  doi       = {10.1109/ICCWorkshops63917.2026.11586446}
}

@misc{Astanjin2026FullPASS,
  author       = {Astanjin, Morteza Barzegar and Azimi-Abarghouyi,
                  Seyed Mohammad and Eriksson, Thomas and Valkama, Mikko},
  title        = {{FullPASS}: Geometry Optimization for Full-Duplex
                  Pinching-Antenna Systems},
  year         = {2026},
  month        = {Jul.},
  eprint       = {2607.19546},
  archivePrefix = {arXiv},
  primaryClass = {cs.IT},
  doi          = {10.48550/arXiv.2607.19546},
  url          = {https://arxiv.org/abs/2607.19546}
}

@book{boyd2004convex,
  author    = {Stephen Boyd and Lieven Vandenberghe},
  title     = {Convex Optimization},
  publisher = {Cambridge University Press},
  year      = {2004},
  address   = {Cambridge},
  isbn      = {9780521833783}
}

@article{li2026pinching,
  title={Pinching Antenna-Assisted Full-Duplex Communication Systems},
  author={Li, Xuan and Lei, Xianfu and Wu, Mingjiang and Tegos, Sotiris A and Diamantoulakis, Panagiotis D and Karagiannidis, George K},
  journal={arXiv preprint arXiv:2609.29200},
  year={2026}
}

@article{johnson2002perturbation,
  title={Perturbation theory for {Maxwell’s} equations with shifting material boundaries},
  author={Johnson, Steven G and Ibanescu, Mihai and Skorobogatiy, MA and Weisberg, Ori and Joannopoulos, JD and Fink, Yoel},
  journal={Physical Review E},
  volume={65},
  number={6},
  pages={066611},
  year={Jun. 2002},
  publisher={APS}}

@book{okamoto2021fundamentals,
  title={Fundamentals of optical waveguides},
  author={Okamoto, Katsunari},
  year={2021},
  publisher={Elsevier}
}

@article{ding2024flexible,
  author={Ding, Zhiguo and Schober, Robert and Vincent Poor, H.},
  journal={IEEE Trans. Commun.}, 
  title={Flexible-Antenna Systems: A Pinching-Antenna Perspective}, 
  year={Mar. 2025},
  volume={73},
  number={10},
  pages={9236-9253},
  doi={10.1109/TCOMM.2025.3555866}}

@book{balanis2012advanced,
  title={Advanced engineering electromagnetics},
  author={Balanis, Constantine A},
  year={2012},
  publisher={John Wiley \& Sons}
}

@INPROCEEDINGS{11591116,
  author={Hu, Shaokang and Zhao, Ruotong and Liao, Yihuan and Ng, Derrick Wing Kwan and Yuan, Jinhong},
  title={Sum-Rate Maximization for Pinching Antenna-Assisted {NOMA} Systems with Multiple Dielectric Waveguides},
  booktitle={Proc. IEEE GLOBECOM Wkshps},
  pages={411--417},
  year={2025},
  doi={10.1109/GCWkshps68340.2025.11591116}
}

@book{snyder1983optical,
  author    = {A. W. Snyder and J. D. Love},
  title     = {Optical Waveguide Theory},
  publisher = {Springer},
  address   = {Boston, MA, USA},
  year      = {1983},
  doi       = {10.1007/978-1-4613-2813-1}
}

@ARTICLE{9738442,
  author={Zhang, Haiyang and Shlezinger, Nir and Guidi, Francesco and Dardari, Davide and Imani, Mohammadreza F. and Eldar, Yonina C.},
  journal=wcom,
  title={Beam Focusing for Near-Field Multiuser {MIMO} Communications}, 
  year={Sep. 2022},
  volume={21},
  number={9},
  pages={7476-7490},
  doi={10.1109/TWC.2022.3158894}}

@book{goldsmith2005wireless,
  title={{Wireless Communications}},
  author={Goldsmith, Andrea},
  year={2005},
  publisher={Cambridge {University} {Press}}
}

@ARTICLE{Shi2011IWMMSE,
  author={Shi, Qingjiang and Razaviyayn, Meisam and Luo, Zhi-Quan and He, Chen},
  journal={IEEE Trans. Signal Process.},
  title={An Iteratively Weighted {MMSE} Approach to Distributed Sum-Utility Maximization for a {MIMO} Interfering Broadcast Channel},
  year={2011},
  month={Oct.},
  volume={59},
  number={9},
  pages={4331--4340},
  doi={10.1109/TSP.2011.2147784}
}

@article{suzuki2022pinching,
  author  = {Fukuda, Atsushi and Yamamoto, Hiroto and Okazaki, Hiroshi
             and Suzuki, Yasunori and Kawai, Kunihiro},
  title   = {Pinching Antenna---Using a Dielectric Waveguide as an Antenna},
  journal = {NTT DOCOMO Technical Journal},
  volume  = {23},
  number  = {3},
  pages   = {5--12},
  month   = jan,
  year    = {2022}
}

@article{yates1995framework,
  author={R. D. Yates},
  journal={IEEE J. Sel. Areas Commun.},
  title={A framework for uplink power control in cellular radio systems},
  year={1995},
  volume={13},
  number={7},
  pages={1341--1347},
  month={Sep.},
  doi={10.1109/49.414651}
}

@ARTICLE{8310563,
  author={Shen, Kaiming and Yu, Wei},
  journal=sip, 
  title={Fractional Programming for Communication Systems—Part {II}: Uplink Scheduling via Matching}, 
  year={Mar. 2018},
  volume={66},
  number={10},
  pages={2631-2644},
  doi={10.1109/TSP.2018.2812748}}

@ARTICLE{9449661,
  author={Yu, Xianghao and Xu, Dongfang and Ng, Derrick Wing Kwan and Schober, Robert},
  journal={IEEE Trans. Commun.}, 
  title={{IRS}-Assisted Green Communication Systems: Provable Convergence and Robust Optimization}, 
  year={Jun. 2021},
  volume={69},
  number={9},
  pages={6313-6329},
  doi={10.1109/TCOMM.2021.3087794}}

@article{Li2026Geometry,
  author= {Haoyang Li and Weidong Liu and Zhensheng Chen and Chaoyun Song and Gaojie Chen},
    journal={IEEE Commun. Lett.}, 
  title={Geometry-Dependent Radiation of Pinching Antennas: Theory, Simulation, and Measurement}, 
  year={May 2026},
  volume={30},
  number={},
  pages={2093-2097},
  doi={10.1109/LCOMM.2026.3696362}}

\end{document}